# Giant Hall Switching by Surface-State-Mediated Spin-Orbit Torque in a Hard Ferromagnetic Topological Insulator

*Lixuan Tai, Haoran He, Su Kong Chong, Huairuo Zhang, Hanshen Huang, Gang Qiu, Yuxing Ren, Yaochen Li, Hung-Yu Yang, Ting-Hsun Yang, Xiang Dong, Bingqian Dai, Tao Qu, Qingyuan Shu, Quanjun Pan, Peng Zhang, Fei Xue, Jie Li, Albert V. Davydov, and Kang L. Wang**

L. Tai, H. He, S. K. Chong, H. Huang, G. Qiu, Y. Ren, Y. Li, H.-Y. Yang, T.-H. Yang, X. Dong, B. Dai, T. Qu, Q. Shu, Q. Pan, P. Zhang, K. L. Wang

University of California, Los Angeles, California 90095, United States

E-mail: wang@ee.ucla.edu

H. Zhang

Theiss Research, Inc., La Jolla, California 92037, United States

H. Zhang, A. V. Davydov

Materials Science and Engineering Division, National Institute of Standards and Technology (NIST), Gaithersburg, Maryland 20899, United States

F. Xue

Department of Physics, University of Alabama at Birmingham, Birmingham, Alabama 35294, United States

J. Li

School of Materials Science and Engineering, Shanghai University, Shanghai 200444, China

Topological insulators (TI) and magnetic topological insulators (MTI) can apply highly efficient spin-orbit torque (SOT) and manipulate the magnetization with their unique topological surface states with ultra-high efficiency. Here, efficient SOT switching of a hard MTI, V-doped $(Bi,Sb)_2Te_3$ (VBST), with a large coercive field that can prevent the influence of an external magnetic field, is demonstrated. A giant switched anomalous Hall resistance of 9.2 kΩ is realized, among the largest of all SOT systems, which makes the Hall channel a good readout and eliminates the need to fabricate complicated magnetic tunnel junction (MTJ) structures. The SOT switching current density can be reduced to $2.8\times10^5$ A cm$^{-2}$, indicating its high efficiency. Moreover, as the Fermi level is moved away from the Dirac point by both gate and composition tuning, VBST exhibits a transition from edge-state-mediated to surface-state-mediated transport, thus enhancing the SOT effective field to $(1.56\pm0.12)\times10^{-6}$ T A$^{-1}$ cm$^2$ and the interfacial charge-to-spin conversion efficiency to $3.9\pm0.3$ nm$^{-1}$. The findings establish VBST as an extraordinary candidate for energy-efficient magnetic memory devices.

## 1. Introduction

Spin-orbit torque (SOT) uses relativistic spin-orbit interaction in heavy elements to convert charge current to spin current and manipulate the magnetization.[1] Heavy elements are usually placed adjacent to magnetic materials and generate a spin current via mechanisms such as spin Hall effect[2–6] or Rashba-Edelstein effect[7–9] to exert spin–orbit torque (SOT) and switch the magnets. As a writing method for magnetoresistive random access memories (MRAM), SOT has great advantages of low writing current and high reliability, so it has promising applications in high-efficiency non-volatile memory.[10–12]

Current-induced SOT switching was first demonstrated in heavy metal/ferromagnet (HMFM) bilayer heterostructures, where the heavy metal acts as a SOT source material.[5,6,13,14] The material choices for SOT sources have been extended to various quantum and 2D materials, such as the Weyl semimetal $WTe_2$.[15,16]

Compared with these conducting materials, topological insulators (TI) are more advantageous as SOT source materials because their insulating bulk and conducting spin-polarized surface states can enable much higher SOT efficiency.[17–20] In TI, the spin-orbit coupling is large enough to invert the band structure and create unique topological surface states (TSS), represented by a linearly dispersed Dirac cone in the band structure. These topological surface states are spin-polarized and have spin-momentum locking, meaning the spin polarization uniquely depends on the momentum.[21–26] Various techniques have been developed to determine the SOT efficiency in TI, including spin-torque ferromagnetic resonance (ST-FMR),[17,20,27,28] spin pumping,[29–31] magneto-optical Kerr effect (MOKE),[32,33] loop shift method in transport,[32,34] and second harmonic measurements in transport.[18,19,33,35,36] The SOT efficiency can be quantified by the spin Hall angle $\theta_{SH}$, which is defined as the ratio between spin current density $J_S$ and electric current density $J_C$, or $\theta_{SH} = J_S / J_C$, and TI's spin Hall angle is reported to be as large as 140 at 1.9 K[18,19] and around 1-3 at room temperature,[28,35,36] although the spin Hall angle only applies to the 3D bulk origin of SOT (like the spin Hall effect) and might not be physically meaningful.[20,30] So far, SOT has been explored in various TI/non-TI magnet bilayers for efficient current-induced magnetization switching.[28,33–40]

Another approach to utilizing the highly efficient SOT of TI is using its magnetic counterparts, the magnetic topological insulators (MTI). MTI is more advantageous than TI/non-TI magnet bilayers because it does not need another heterogeneous magnetic layer for magnetization switching and is thus free from problems with the bilayer interface and shunting in the metallic ferromagnetic layer. By doping TI with magnetic ions like Cr, Mn, or V, ferromagnetism can be established within TI itself. A gap can be opened at the crossing point of the Dirac cone, the Dirac point, leaving pure edge state transport only and thus hosting the quantum anomalous Hall effect.[41–45] Therefore, MTI or TI/MTI bilayer has reached one of the highest SOT efficiency ever reported.[18,19,32,46]

Among all the various MTIs, V-doped $(Bi,Sb)_2Te_3$ (VBST) is unique because it is a hard ferromagnetic topological insulator with a larger coercive field (>0.5 T, one order of magnitude larger than its Cr-doped counterparts), thus making it robust against the influences of external magnetic fields. Thanks to its robust ferromagnetism, it also exhibits spontaneous magnetization (and quantum anomalous Hall effect) without the need for magnetic field training or field cooling, in contrast to its Cr-doped counterparts.[45] All these unique properties of VBST as a hard ferromagnetic topological insulator make it an attractive candidate for spintronics applications.

Here, we demonstrate efficient current-driven magnetization switching of a hard ferromagnetic topological insulator, V-doped $(Bi,Sb)_2Te_3$ (VBST) by its surface-state-mediated SOT. A giant switched anomalous Hall resistance, a low switching current density, or a high switching ratio can be realized. By analyzing the second-harmonic signals of the anomalous Hall resistance, we calibrate the efficiency of the SOT, showing a high effective field to current ratio, which is three orders of magnitude larger than those reported in HMFMs. More interestingly, Fermi level dependence by tuning the gate voltage or the material composition reveals that as the surface state transport replaces edge state transport, more carriers from the surface states significantly enhance SOT. This hard ferromagnetism, giant Hall switching, and efficient surface-state-mediated SOT in VBST provide a perfect platform for energy-efficient magnetic memory devices.

## 2. Material Characterizations and Device Structure

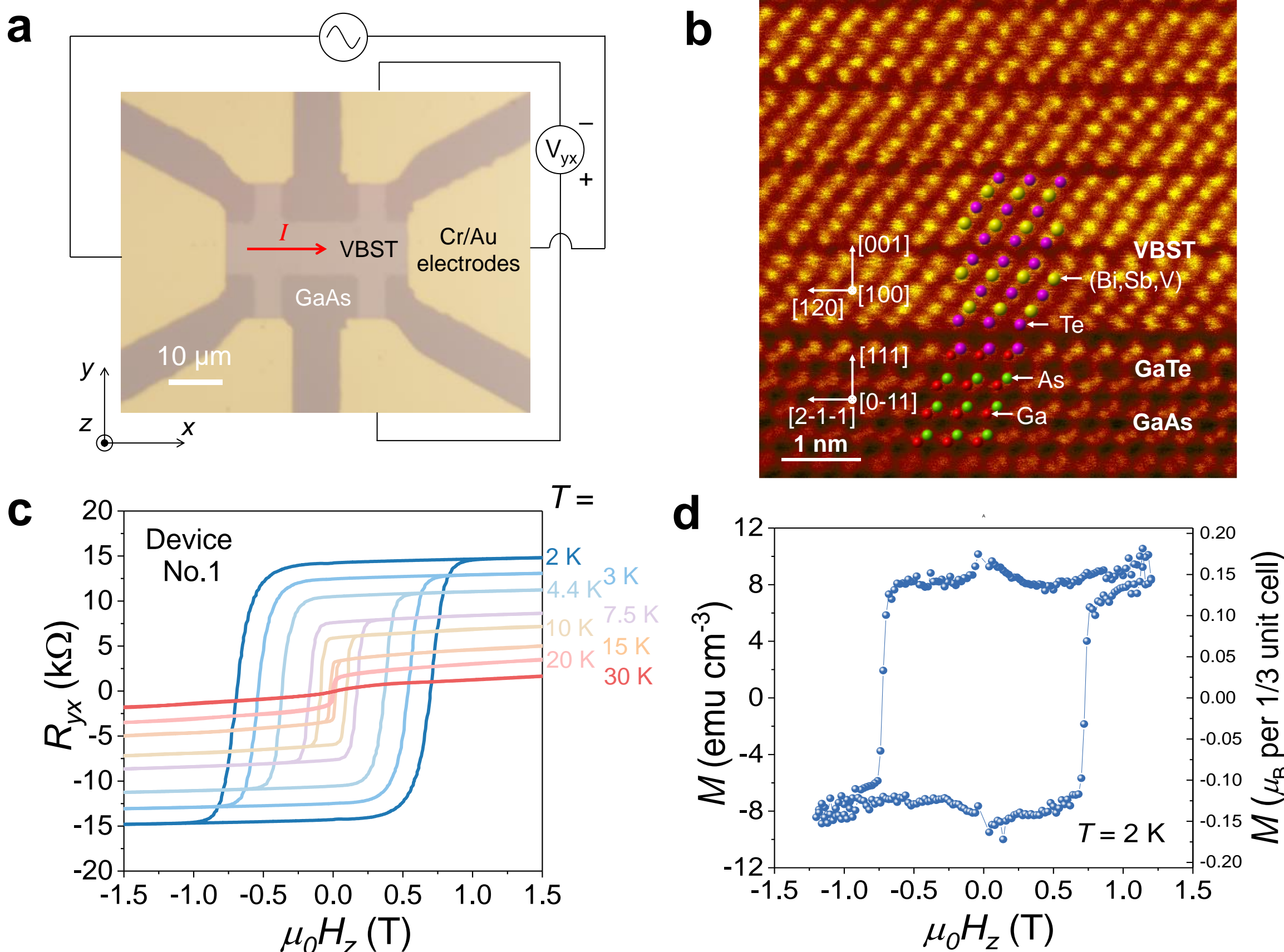

**Figure 1.** Experimental set-up and the magnetic properties of the V-doped $(Bi,Sb)_2Te_3$ (VBST) films. a) Optical micrograph of the Hall bar device. b) A cross-sectional HAADF-STEM image of the VBST film on the GaAs substrate. c) Anomalous Hall resistance $R_{yx}$ as a function of an external out-of-plane magnetic field $H_z$ at various temperatures in Device No.1. d) Magnetization $M$ as a function of an external out-of-plane magnetic field $H_z$ at $T$=2 K.

**Figure 1**a shows the Hall bar device structure with schematic illustrations of the experimental set-up, where the bright region shows the Hall bar of the V-doped $(Bi,Sb)_2Te_3$ (VBST) thin film, the dark region shows the semi-insulating GaAs (111)B substrate and the gold region is the Cr/Au electrodes. The length and width of the Hall bar are 20 μm and 10 μm, respectively. All devices (Device No.1-8) in this work have the same Hall bar geometry, and only Device No.3 has an additional top gate.

In order to study the crystalline quality of the thin film, cross-sectional atomic-resolution HAADF-STEM (high-angle annular dark-field scanning transmission electron microscopy) was carried out. The captured cross-sectional image of the VBST sample, as presented in Figure 1b, shows that the quintuple layer (QL) structure of the epitaxial rhombic VBST film and van der Waals (vdW) gaps between every two QLs, indicating high crystalline quality. The image also reveals an atomically sharp interface between the VBST and the Te-terminated GaAs (111)B substrate from the Te-rich pre-annealing process to remove the top oxide layer. In conclusion, the cross-sectional STEM image confirms the high crystalline quality of VBST.

For the transport properties of VBST samples, the Hall resistance is presented in Figure 1c as a function of an external out-of-plane magnetic field $H_z$ at various temperatures on a 6 QL $V_{0.14}(Bi_{0.27}Sb_{0.73})_{1.86}Te_3$ Hall bar (Device No.1). A wide hysteresis loop from the anomalous Hall effect (AHE) with a large coercive field of 0.7 T is observed at $T$=2 K, one order of magnitude larger than its Cr-doped counterparts, indicating the nature of VBST as a hard ferromagnet. The sample also exhibits a large anomalous Hall resistance of 14.8 kΩ at $T$=2 K. The anomalous Hall resistance and coercive field shrinks at higher temperatures, and the AHE persists up to 30 K.

In Figure 1d, the magnetization $M$ of VBST as a function of an external out-of-plane magnetic field $H_z$ at $T$=2 K, from which the diamagnetic background has been removed, also reveal a large coercive field of 0.7 T and a small saturation magnetization of $M_S$=8.3 emu cm$^{-3}$, which corresponds to a net magnetic moment of $0.15\mu_B$ ($\mu_B$ is Bohr magneton) per 1/3 unit cell (5 atoms of $(Bi,Sb)_2Te_3$ included).

VBST's unique magnetic properties make it an ideal candidate for spintronics materials. Its large coercive field, which is more than one order of magnitude larger than typical ferromagnets, can prevent the influence of an external magnetic field and make the magnetic storage robust against external magnetic perturbations. Its small saturation magnetization, which is two orders of magnitude smaller than typical ferromagnets, makes the stray field small and eliminates unwanted magnetic crosstalk between neighboring devices.

## 3. Current-Induced Spin-Orbit Torque Switching

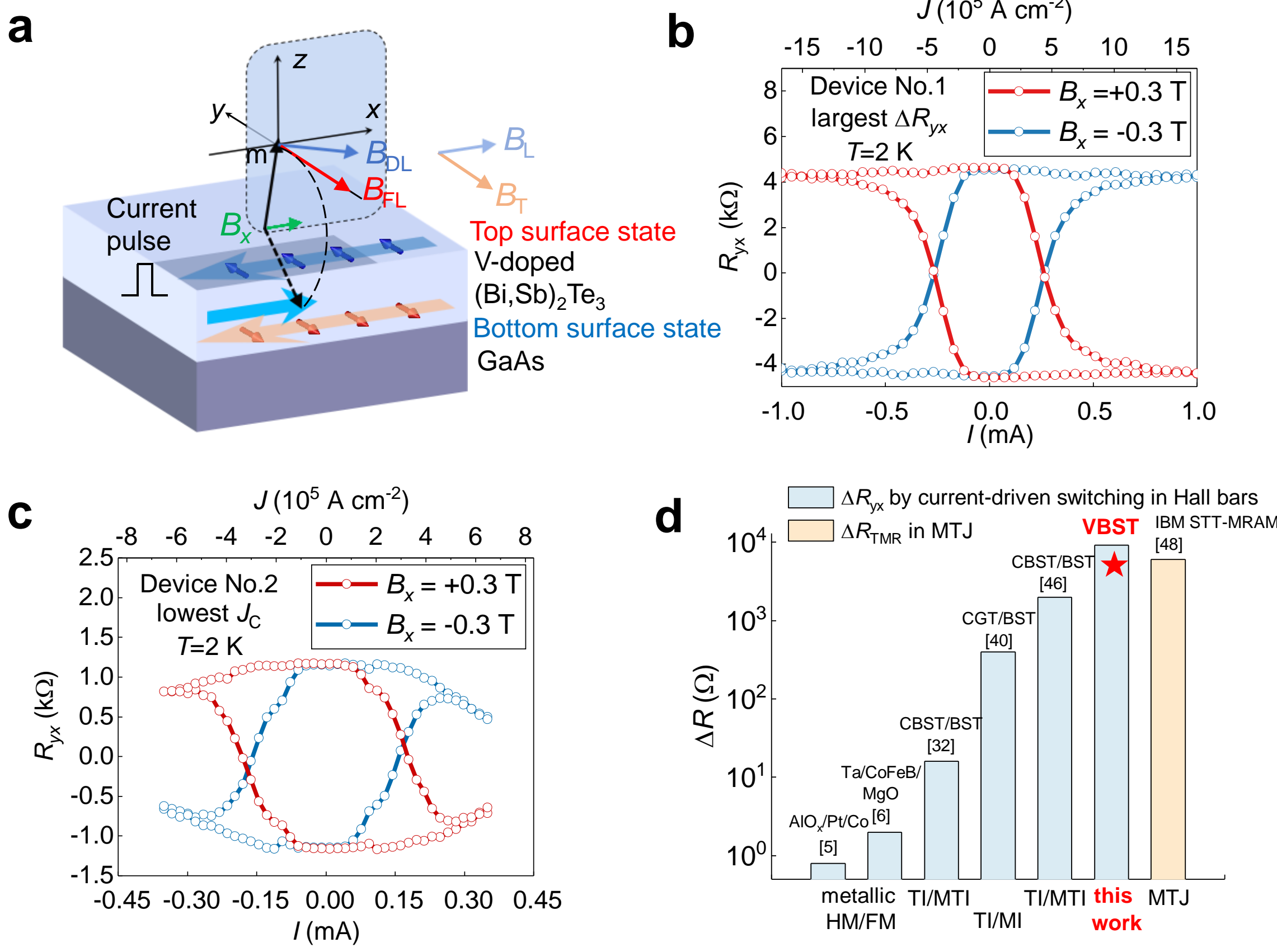

**Figure 2.** Current-induced switching of magnetization in VBST. a) Schematic illustration of current-induced magnetization switching with damping-like (DL) and field-like (FL) SOT effective fields $B_{DL}$ and $B_{FL}$. $B_L$ or $B_T$ is an applied external longitudinal or transverse magnetic field for Figure 3. b, c) The Hall resistance $R_{yx}$ after applying each writing pulse as a function of the pulse amplitude $I$ as well as the current density $J$, taken in (b) Device No.1 and (c) Device No.2, with an in-plane assisting magnetic field $B_x$ of ±0.3 T. All measurements were carried out at $T$=2 K. d) Comparison between the switched resistances by current in various schemes.

Current-driven SOT switching of magnetization was performed in the hard ferromagnetic topological insulator VBST. **Figure 2**a shows a schematic illustration of surface-state-mediated SOT switching in VBST. When a current is injected into the $+\mathbf{x}$ direction, electrons will move into the $-\mathbf{x}$ direction through the top and bottom surface states that carry $+\mathbf{y}$ and $-\mathbf{y}$ spin polarizations, respectively, from the spin-momentum locking mechanism of topological insulators. Although the top and bottom surface states carry opposite spin directions so their contributions to the SOT cancel each other, the structural asymmetry in the VBST/GaAs structure can induce different interfacial properties and different top and bottom surface state carrier densities, thus leading to net current-induced SOT.[19,47] The spin accumulation exerts torques on the magnetization $\mathbf{m}$, thus leading to its flipping. The SOT can be decomposed into two parts, namely the field-like (FL) and damping-like (DL) torques, as in **Equation 1**.[1]

$$\mathbf{T} = \mathbf{M} \times (\mathbf{B}_{\mathrm{FL}} + \mathbf{B}_{\mathrm{DL}}) = -M_{\mathrm{S}}\,[B_{\mathrm{FL}}\mathbf{m} \times \boldsymbol{\zeta} + B_{\mathrm{DL}}\mathbf{m} \times (\mathbf{m} \times \boldsymbol{\zeta})] \quad (1)$$

Here, $\mathbf{B}_{\mathrm{FL}} = B_{\mathrm{FL}}\boldsymbol{\zeta}$ and $\mathbf{B}_{\mathrm{DL}} = B_{\mathrm{DL}}\mathbf{m} \times \boldsymbol{\zeta}$ represent the FL and DL effective fields, $\mathbf{m} = \mathbf{M}/M_{\mathrm{S}}$ is the magnetization unit vector, $\mathbf{M}$ is the magnetization, $M_{\mathrm{S}}$ is the saturation magnetization, and $\boldsymbol{\zeta}$ is the unit vector of the net SOT spin polarization. The FL torque produces the precession of the magnetization, and the DL torque produces the damping of the magnetization towards its equilibrium state, thus mainly contributing to the switching. Therefore, the DL torque is usually used to evaluate the SOT efficiency. For deterministic switching, an in-plane external assisting magnetic field $B_x$ is needed to drive $\mathbf{m}$ towards $\mathbf{x}$ direction and gives $\mathbf{B}_{\mathrm{DL}}$ a necessary $\mathbf{z}$ component to flip the magnetization.

The measurement scheme of current-induced switching is by applying a series of current pulses and outlined in Section 5 of the Supporting Information. Figure 2b presents the Hall resistance $R_{yx}$ of Device No.1, 6 QL $V_{0.14}(Bi_{0.27}Sb_{0.73})_{1.86}Te_3$ as a function of the writing pulse amplitude $I$ and the current density $J$, when $B_x = \pm 0.3$ T. All measurements were carried out at the ambient temperature of $T$=2 K. When a current pulse is applied in the $+\mathbf{x}$ direction with an in-plane external assisting magnetic field $B_x$ driving $\mathbf{m}$ towards $+\mathbf{x}$ direction, it gives $\mathbf{B}_{\mathrm{DL}} = B_{\mathrm{DL}}\mathbf{m} \times \boldsymbol{\zeta}$ a $-\mathbf{z}$ component to flip the magnetization from $+\mathbf{z}$ to $-\mathbf{z}$. This is only possible with $\boldsymbol{\zeta} = -\mathbf{y}$, which is the spin direction of the bottom surface state. Therefore, from the switching polarity, it is determined that the bottom surface states dominate the switching process.

Notably, the difference in the anomalous Hall resistance $\Delta R_{yx}$ switched by the current is as large as 9.2 kΩ, among the largest of all SOT systems. This is thanks to the large anomalous Hall resistance of 14.8 kΩ that Device No.1 exhibits. The switching current density $J_{\mathrm{C}}$ is also as low as $4.5 \times 10^5$ A cm$^{-2}$, and the switching ratio of the magnetic domains is about 31%.

To further reduce the switching current density, Device No.2, 6 QL $V_{0.14}(Bi_{0.18}Sb_{0.82})_{1.86}Te_3$ with a higher interfacial charge-to-spin conversion efficiency of 3.9 nm$^{-1}$ (than Device No.1's 2.0 nm$^{-1}$, which we will discuss later) is used. From Figure 2c, Device No.2 shows a much lower switching current density of $2.8 \times 10^5$ A cm$^{-2}$ than $4.5 \times 10^5$ A cm$^{-2}$ in Device No.1, which is almost 40% reduction and consistent with the increase of SOT efficiency. The switching current density is two orders of magnitude smaller than typical HMFM bilayers,[1,5,6,13,14] and on the same order of magnitude but slightly larger than that in Cr-doped $(Bi,Sb)_2Te_3$ (CBST) ($\sim 2.57 \times 10^5$ A cm$^{-2}$) under the same pulse-driven switching scheme,[32] because VBST's large perpendicular magnetic anisotropy (PMA) as a hard ferromagnet creates an additional energy barrier for magnetization switching.

The switching ratio of the magnetic domains is not 100% because of inhomogeneous magnetic domains, so some earlier works on the switching of Cr-doped MTI show a switching ratio of <10%.[18,19,32] By enhancing the PMA of VBST with increased V doping, the switching ratio can be further improved to 60% (see Figure S9 in the Supporting Information).

To demonstrate how giant the switching of the Hall resistance in VBST is, a comparison between the switched resistances by current in various schemes is presented in Figure 2d. The changes in the anomalous Hall resistance by current-driven SOT switching are presented in HMFM systems including $AlO_x$/Pt/Co & Ta/CoFeB/MgO,[5,6] TI/MTI systems like Cr-doped $(Bi,Sb)_2Te_3$/$(Bi,Sb)_2Te_3$ (CBST/BST),[32,46] TI/magnetic insulator (MI) systems like $Cr_2Ge_2Te_6$/$(Bi,Sb)_2Te_3$ (CGT/BST),[40] and V-doped $(Bi,Sb)_2Te_3$ (VBST, this work). For comparison, the change in the tunnel magnetoresistance (TMR) of the magnetic tunnel junction (MTJ) in IBM's spin-transfer torque (STT) MRAM is also listed here.[48] From the comparison, the giant switching of Hall resistance of 9.2 kΩ in VBST is almost four orders of magnitude higher than those in conventional HMFM systems[5,6] or TI/metallic FM systems (~1 Ω),[34–36] making it among the largest of all SOT systems. The change of anomalous Hall resistance is even comparable to or greater than the change of TMR in MTJ devices, which makes the Hall channel a good readout and eliminates the need to fabricate complicated three-terminal SOT-MTJ structures.

In summary, the current-driven switching tests further establish VBST as an ideal spintronics material for its giant switched anomalous Hall resistance, larger switching ratio, and low critical switching current density due to its unique band topology, large anomalous Hall resistance, and highly efficient surface-state-mediated SOT.

## 4. Second-Harmonic Measurements of SOT Effective Fields

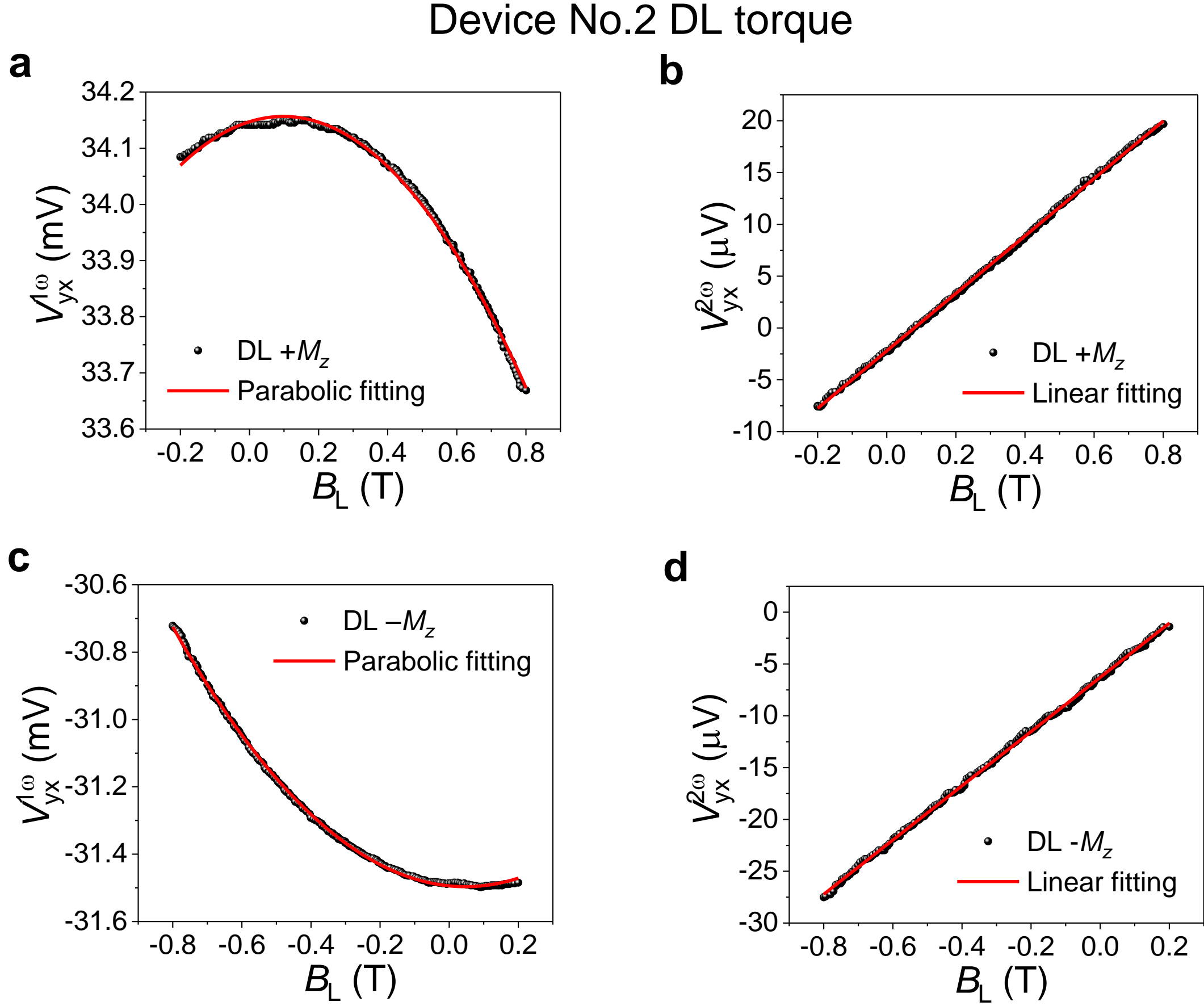

**Figure 3.** In-plane low-field second-harmonic measurements of the damping-like (DL) SOT effective field for Device No.2. a) In-phase first-harmonic and b) out-of-phase second-harmonic Hall voltages $V_{yx}^{1\omega}$ and $V_{yx}^{2\omega}$ as a function of an external longitudinal magnetic field $B_L$ when the magnetization is along the $+\mathbf{z}$ direction ($+M_z$). c) In-phase first-harmonic and d) out-of-phase second-harmonic Hall voltages when the magnetization is along the $-\mathbf{z}$ direction ($-M_z$). All measurements were carried out at $T$=2 K by applying a sinusoidal current of 10 μA.

In order to determine either the damping-like (DL) or the field-like (FL) SOT effective field $B_{DL}$ and $B_{FL}$ and thus the SOT efficiency in VBST, in-plane low-field second-harmonic measurements were conducted by applying either an external longitudinal or transverse magnetic field, $B_L$ or $B_T$ to slightly tilt the magnetization, as illustrated in Figure 2a. The first and second harmonic Hall voltages $V_{yx}^{1\omega}$ and $V_{yx}^{2\omega}$ were measured by lock-in amplifiers at the ambient temperature of $T$=2 K with a sinusoidal excitation current of 10 μA (root mean square value). **Figure 3** presents one of the results of the $V_{yx}^{1\omega}$ and $V_{yx}^{2\omega}$ as a function of $B_L$ in Device No.2, 6 QL $V_{0.14}(Bi_{0.18}Sb_{0.82})_{1.86}Te_3$ when the magnetization was initialized to either $+\mathbf{z}$ or $-\mathbf{z}$ direction, revealing a parabolic dependence of $V_{yx}^{1\omega}$ and a linear dependence of $V_{yx}^{2\omega}$ on $B_L$, respectively. We can obtain the DL effective field $B_{DL}$ from the data by **Equation 2**.[49,50]

$$B_{\mathrm{DL}} = -2\frac{\partial V_{yx}^{2\omega}}{\partial B_{\mathrm{L}}} \Big/ \frac{\partial^2 V_{yx}^{1\omega}}{\partial B_{\mathrm{L}}^2} \tag{2}$$

By using the above equation, $B_{\mathrm{DL}} = (2.59 \pm 0.20) \times 10^{-2}$ T. As an important indicator of the SOT efficiency, the DL SOT effective field to electric current density ratio is thus $B_{\mathrm{DL}}/J = (1.56 \pm 0.12) \times 10^{-6}\ \mathrm{T\ A^{-1}\ cm^2}$, more than three orders of magnitude larger than typical HMFM bilayer heterostructures.[1,5,6,13,14] Here, the device parameters of a width of 10 μm and a thickness of 6 nm are used.

Another important indicator of SOT efficiency is the interfacial charge-to-spin conversion efficiency $q_{\mathrm{ICS}}$, which is defined as the ratio of the 3D spin current density $J_{\mathrm{S,3D}}$ to the 2D electric current density $J_{\mathrm{C,2D}}$ as in **Equation 3**. This indicator applies to the 2D interfacial origin of SOT, like the Rashba-Edelstein Effect and topological surface states.[6,20]

$$q_{\mathrm{ICS}} = \frac{J_{\mathrm{S,3D}}}{J_{\mathrm{C,2D}}} = \frac{2eM_{\mathrm{S}}B_{\mathrm{DL}}}{\hbar J_{\mathrm{C,3D}}} \tag{3}$$

Here, $M_{\mathrm{S}}$ is the saturation magnetization and $J_{\mathrm{C,3D}}$ is the 3D electric current density. This is a better metric than the spin Hall angle $\theta_{\mathrm{SH}} = J_{\mathrm{S,3D}}/J_{\mathrm{C,3D}} = q_{\mathrm{ICS}}t$, where $t$ is the thickness, since it only applies to the 3D bulk origin of SOT (like the spin Hall effect) and might not be physically meaningful in MTI. The $q_{\mathrm{ICS}}$ of VBST in Device No.2 is as large as 3.9±0.3 nm$^{-1}$, indicating its high efficiency.[1,5,6,13,14]

Note that although this in-plane low-field second-harmonic method is subject to the artificial contributions from thermoelectric effects, such as the Nernst effect, such contributions are proved to be negligible by a different method of the angle-resolved second harmonic measurements in Section 11 of the Supporting Information.[51] This in-plane low-field second-harmonic method also has a negligible contribution from the asymmetric magnon scattering reported by Yasuda *et al.*[46] because in order to have a sufficiently large contribution from the asymmetric magnon scattering, the magnetization has to be driven almost fully in plane by a large external field. However, in our method, with the large perpendicular magnetic anisotropy (PMA) of VBST and a small external field, only up to 1%-2% of the magnetization is driven in plane.

In short, the second-harmonic measurements further demonstrate the huge potential of VBST as an ideal spintronics material for its ultra-high SOT efficiency, which arises from the spin-momentum-locked surface state and is free from interfacial or shunting problems in bilayers.

## 5. Fermi Level Dependence of SOT by Gate and Composition Tuning

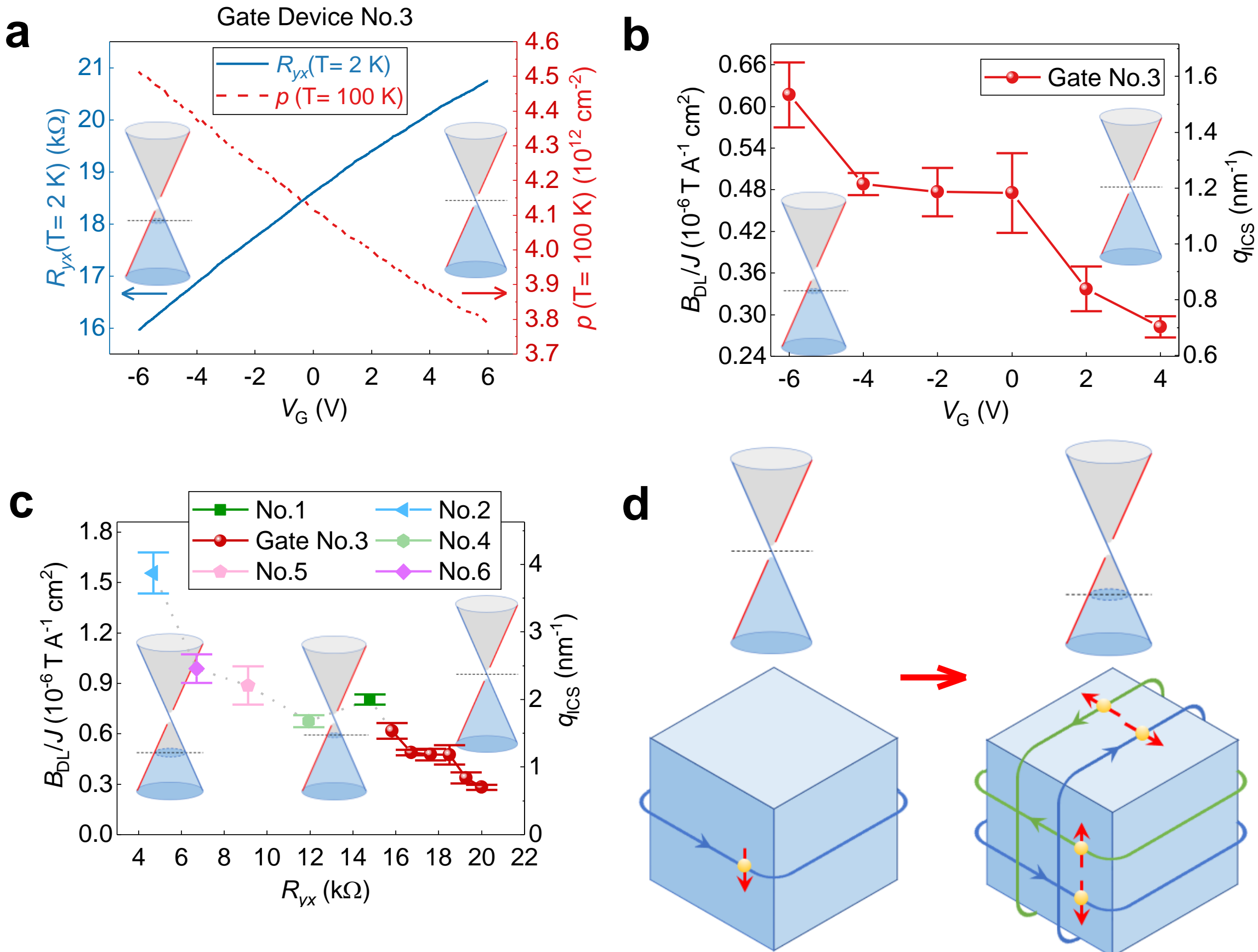

**Figure 4.** Electric field tuning and Fermi level dependence of VBST SOT. a) Gate dependence of anomalous Hall resistance $R_{yx}$ at $T$=2 K and hole density $p$ at 100 K in Device No.3. b) Gate dependence of the SOT efficiency, as indicated by the damping-like effective field to current density ratio $B_{DL}/J$ and the interfacial charge-to-spin conversion efficiency $q_{ICS}$, in Device No.3. c) The dependence of the SOT efficiency on Fermi level, as indicated by the anomalous Hall resistance $R_{yx}$ at 2 K in Device No.1-6. The error bars in (b) and (c) represent the standard deviations from the measurements. d) A schematic showing the transition from the edge-state-dominated transport (when the Fermi level is in the exchange gap of the Dirac cone) to surface-state-dominated transport (when the Fermi level is away from the Dirac point).

### 5.1. Gate Tuning

In order to determine the SOT's dependence on Fermi level, 6 QL $V_{0.14}(Bi_{0.29}Sb_{0.71})_{1.86}Te_3$ was fabricated into a Hall bar with a top gate (Device No.3). Notably, this device exhibits the quantum anomalous Hall effect at $T$=100 mK (see Figure S3 in the Supporting Information) with chiral edge states in transport only, indicating that the Fermi level is within the exchange gap of the Dirac cone. By tuning the top gate voltage $V_G$ from +6 V to -6 V, the anomalous Hall resistance $R_{yx}$ in **Figure 4**a shrinks from 20.8 kΩ to 16.0 kΩ at $T$=2 K as the 2D hole density $p$ increases from $3.79\times10^{12}$ cm$^{-2}$ to $4.51\times10^{12}$ cm$^{-2}$ at $T$=100 K, indicating Fermi level is tuned lower and away from the Dirac point.[41,45] Here, 100 K is chosen for carrier density because only the ordinary Hall effect exists, and at 2 K, a large AHE makes the slope extraction inaccurate. Correspondingly in Figure 4b, the damping-like effective field to current density ratio $B_{DL}/J$ increases from $(2.8\pm0.2)\times10^{-7}$ T A$^{-1}$ cm$^2$ to $(6.2\pm0.5)\times10^{-7}$ T A$^{-1}$ cm$^2$, and so does the interfacial charge-to-spin conversion efficiency $q_{ICS}$ from 0.70±0.03 nm$^{-1}$ to 1.5±0.1 nm$^{-1}$, indicating an increase of the SOT efficiency.

### 5.2. Composition Tuning

**Table 1.** The composition of various devices.

| Device No. | Bi level $x$ | $R_{yx}$(T=2 K) [kΩ] | $B_{DL}/J$ [$10^{-6}$ T A$^{-1}$ cm$^2$] | $q_{ICS}$ [nm$^{-1}$] |
|---|---|---|---|---|
| 1 | 0.27 | 14.8 | 0.80±0.03 | 2.0±0.1 |
| 2 | 0.18 | 4.69 | 1.56±0.12 | 3.9±0.3 |
| 3 | 0.29 | 16.0 to 20.8 | 0.62±0.05 to 0.28±0.02 | 1.5±0.1 to 0.70±0.03 |
| 4 | 0.24 | 11.95 | 0.68±0.04 | 1.7±0.1 |
| 5 | 0.22 | 9.11 | 0.89±0.11 | 2.2±0.3 |
| 6 | 0.20 | 6.70 | 0.99±0.08 | 2.5±0.2 |

In order to further tune the Fermi level beyond the range of electrostatic gating, more Hall bar devices of 6 QL $V_{0.14}(Bi_xSb_{1-x})_{1.86}Te_3$ (Device No.1, 2, 4-6) were fabricated with a lower Bi level $x$ than No.3 so that the Fermi level is tuned even lower and crosses the Dirac cone.[52] The details of each device are listed in **Table 1**, and the doping dependence of anomalous Hall resistance and hole density is presented in Figure S5 in the Supporting Information. To compare with the gate tuning data, the anomalous Hall resistance $R_{yx}$ at $T$=2 K is used as a unifying indicator of the Fermi level because a smaller $R_{yx}$ means a lower Fermi level.[41,45] As the $R_{yx}$ is tuned to 4.69 kΩ in Figure 4c, the $B_{DL}/J$ further increases to $(1.56\pm0.12)\times10^{-6}$ T A$^{-1}$ cm$^2$, and so does the $q_{ICS}$ to 3.9±0.3 nm$^{-1}$ in Device No.2, indicating an enhanced efficiency at lower Fermi level.

Both gate and composition tuning in VBST indicate that the SOT efficiency is significantly enhanced as the device transitions from the edge-state-dominated transport regime (with the Fermi level in the exchange gap) to the surface-state-dominated transport regime (with the Fermi level crossing the Dirac cone), as pictorially illustrated in Figure 4d. This confirms that the SOT in VBST is mediated by the top and bottom surface states. This is because the spin polarization of the edge state $\boldsymbol{\zeta} \parallel \mathbf{m}$ when $\mathbf{m}$ is aligned to the $z$ direction, thus yielding $\mathbf{B}_{\mathrm{DL}} = B_{\mathrm{DL}}\mathbf{m} \times \boldsymbol{\zeta} = \mathbf{0}$, while the spin polarization of the top and bottom surface states $\boldsymbol{\zeta} \perp \mathbf{m}$, thus leading to a non-zero $\mathbf{B}_{\mathrm{DL}}$.

Notably, this work's exploration in MTI's SOT during the transition from edge-state-dominated to surface-state-dominated transport regime is unique, since previous works of the gate tuning of SOT effective field in MTI or MTI/TI bilayer have only explored non-quantized surface-state-dominated transport regimes.[18,19,32,46]

## 6. Conclusion

In conclusion, we have explored SOT in a hard ferromagnetic topological insulator thin film, V-doped $(Bi,Sb)_2Te_3$ (VBST). This material's large coercive field (>0.5 T) makes it robust against external magnetic disturbances, and its small magnetization (~8.3 emu $cm^{-3}$) can minimize the stray field. We have demonstrated efficient SOT switching with a giant switched anomalous Hall resistance of 9.2 kΩ, among the largest of all SOT systems, thus making the Hall channel a good readout. The SOT switching current density can be as low as $2.8\times10^5$ A $cm^{-2}$, indicating ultra-high efficiency, and the switching ratio can also be enhanced to 60%. By second-harmonic analysis, the SOT effective field to electric current density ratio can be as large as $(1.56\pm0.12)\times10^{-6}$ T $A^{-1}$ $cm^2$, and the interfacial charge-to-spin conversion efficiency can be as large as $3.9\pm0.3$ $nm^{-1}$, more than two orders of magnitude larger than those reported in heavy metal/ferromagnet (HMFM) bilayers. Fermi level dependence by both gate tuning and composition tuning indicates that the SOT in VBST is surface-state-mediated, so by tuning the Fermi level away from the Dirac point, more carriers from the surface states tend to enhance the SOT efficiency. This study confirms the potential of VBST as an ideal material candidate for the applications of energy-efficient magnetic memory devices, especially for cryogenic peripheral in-memory computing devices for quantum computing, which also requires low temperatures.[53]

## 7. Experimental Section

*Growth of Materials*: The V-doped $(Bi,Sb)_2Te_3$ (VBST) materials in this paper were grown on epi-ready semi-insulating GaAs (111)B substrates in an ultra-high vacuum, Perkin-Elmer molecular beam epitaxy (MBE) system. Before growth, the substrates were loaded into the MBE chamber and pre-annealed at the temperature of 630 °C in a Te-rich environment to remove the oxide on the surface. During growth, high-purity Bi, Sb, and Te were evaporated from standard Knudsen cells, and V was evaporated from the Thermionics e-Gun™ electron beam evaporator. The substrate was kept at 180 °C. The reflection high-energy electron diffraction (RHEED) *in situ* was used to monitor the quality and thickness of the materials.

*Material Characterizations*: The HAADF-STEM characterization was performed on a 29 QL $V_{0.14}(Bi_{0.22}Sb_{0.78})_{1.86}Te_3$ thin film on a GaAs(111)B substrate with FEI Nova NanoLab 600 DualBeam (SEM/FIB). 0.5 mm Pt was first capped on top of the sample by electron beam-induced deposition to protect its surface. Then 1 μm Pt was deposited by ion beam-induced deposition. Finally, the sample was cleaned by 2 kV Ga ions with a low beam current of 29 pA and a small incident angle of 3 degrees to minimize the damage. An FEI Titan 80-300 probe-corrected STEM/TEM microscope operating at 300 keV was used to capture HAADF-STEM images with atomic resolution.

The magnetization data were also taken on the same sample of 29 QL $V_{0.14}(Bi_{0.22}Sb_{0.78})_{1.86}Te_3$ by a superconducting quantum interference device (SQUID) magnetometer under VSM (vibrating sample magnetometer) mode and an out-of-plane geometry. The SQUID model is Quantum Design MPMS3 with a 7 T superconducting magnet, a base temperature of 1.8 K, and a sensitivity of $1\times10^{-8}$ emu.

*Device Fabrication*: For all the devices (No.1-8), the VBST thin films were fabricated into a 20 μm (length) × 10 μm (width) Hall bar geometry by a standard photolithography process. Cr/Au contact electrodes with thickness of 10/100 nm were deposited using an electron beam evaporator. For Device No.3 only, an additional top gate was fabricated by exfoliating and transferring mica and graphite thin flakes on top to serve as the gate dielectric and the electrode, respectively.

*Transport Measurements*: Low-temperature magneto-transport measurements and current-induced switching tests were performed in a Quantum Design physical property measurement system (PPMS) with a 9 T superconducting magnet and a base temperature of 1.9 K. For magneto-transport measurements, a Keithley 6221 current source was used to generate a source AC current, and multiple lock-in amplifiers (Stanford Research SR830) were used to obtain the first and second-harmonic Hall voltages. For current-induced switching, a Keithley 2636 source meter was used to generate a series of writing and reading current pulses and to obtain the Hall voltages. Gate voltage was applied to the gate electrode of Device No.3 and swept by a Keithley 2636 source meter.

**Supporting Information**

Supporting Information is available from the Wiley Online Library or from the author.

**Author Contributions**

L.T., H.H., and S.K.C. contributed equally to this work. L.T., H.H., and K.L.W. conceived and designed the experiments. K.L.W. supervised the work. L.T., H.-Y.Y., T.-H.Y., X.D., and Y.R. grew the sample. H.Z. and A.V.D. performed the transmission electron microscopy measurement. L.T. performed the SQUID and XRD measurements. S.K.C. and Y.L. fabricated the Hall bar devices. L.T., H.H., and G.Q. carried out the transport measurements. L.T. processed all the data. All authors contributed to the analyses. L.T. and K.L.W. wrote the manuscript with contributions from all authors.

**Acknowledgements**

The authors acknowledge the support from the National Science Foundation (NSF) (DMR-1411085 and DMR-1810163) and the Army Research Office Multidisciplinary University Research Initiative (MURI) under grant numbers W911NF16-1-0472 and W911NF-19-S-0008. In addition, H.Z. acknowledges support from the U.S. Department of Commerce, NIST under financial assistance award 70NANB19H138. A.V.D. acknowledges support from the Material Genome Initiative funding allocated to NIST. F.X. acknowledges the support of the National Science Foundation under Grant No. OIA-2229498 and Oak Ridge Associated Universities Ralph E. Powe Junior Faculty Enhancement Award. J.L. acknowledges the support of the "Leading Talent" program of Shanghai and the National Natural Science Foundation of China (No. 12304089).

**Conflict of Interest Statement**

The authors declare no conflict of interest.

**Data Availability Statement**

The data that support the findings of this study are available from the corresponding author upon reasonable request.

Supporting Information

**Giant Hall Switching by Surface-State-Mediated Spin-Orbit Torque in a Hard Ferromagnetic Topological Insulator**

*Lixuan Tai, Haoran He, Su Kong Chong, Huairuo Zhang, Hanshen Huang, Gang Qiu, Yuxing Ren, Yaochen Li, Hung-Yu Yang, Ting-Hsun Yang, Xiang Dong, Bingqian Dai, Tao Qu, Qingyuan Shu, Quanjun Pan, Peng Zhang, Fei Xue, Jie Li, Albert V. Davydov, and Kang L. Wang**

## 1. RHEED Pattern

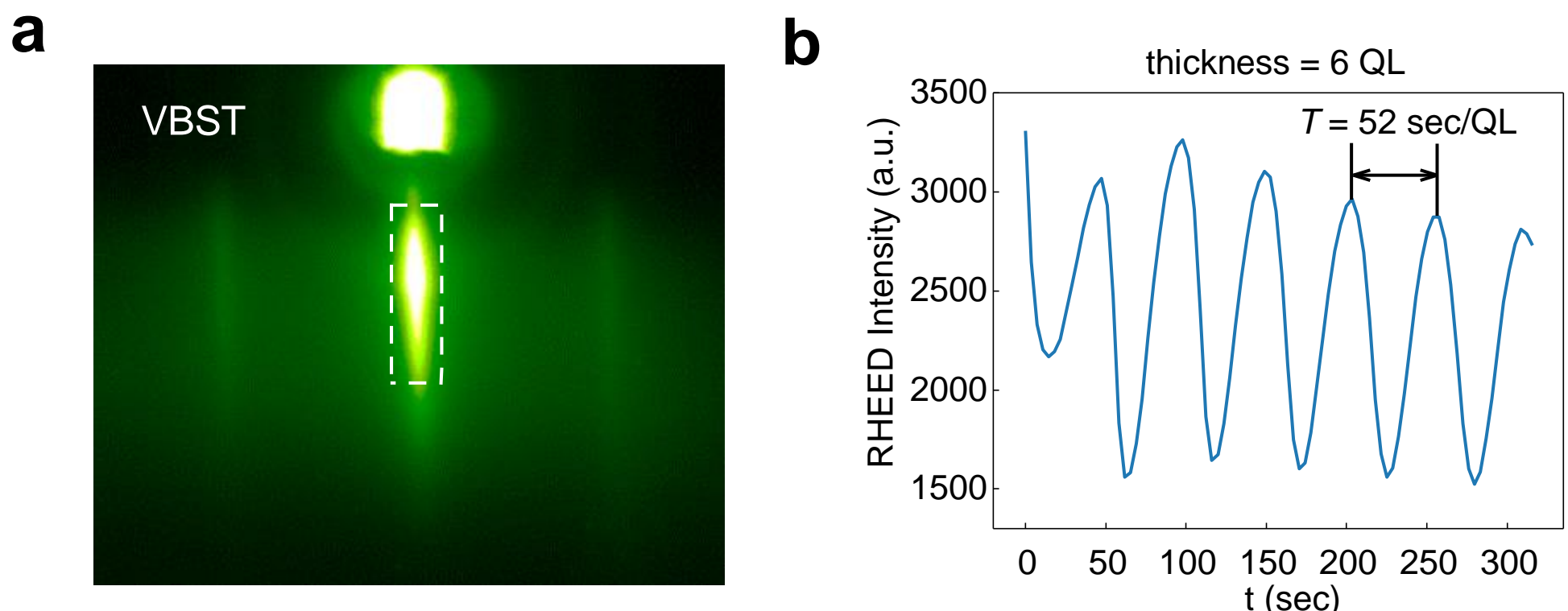

**Figure S1.** a) RHEED pattern of VBST b) RHEED intensity inside the square as a function of growth time.

The growth process of VBST was monitored by the reflection high-energy electron diffraction (RHEED) *in situ*. The sharp and streaky lines in the RHEED pattern, as shown in **Figure S1**a, indicate good quality of the epitaxial single crystal. The RHEED intensity also shows nice oscillation with the growth time in Figure S1b, with every peak indicating the growth of one additional QL. The growth period is 52 sec for every QL, and the thickness is 6 QL for this sample.

## 2. XRD Pattern

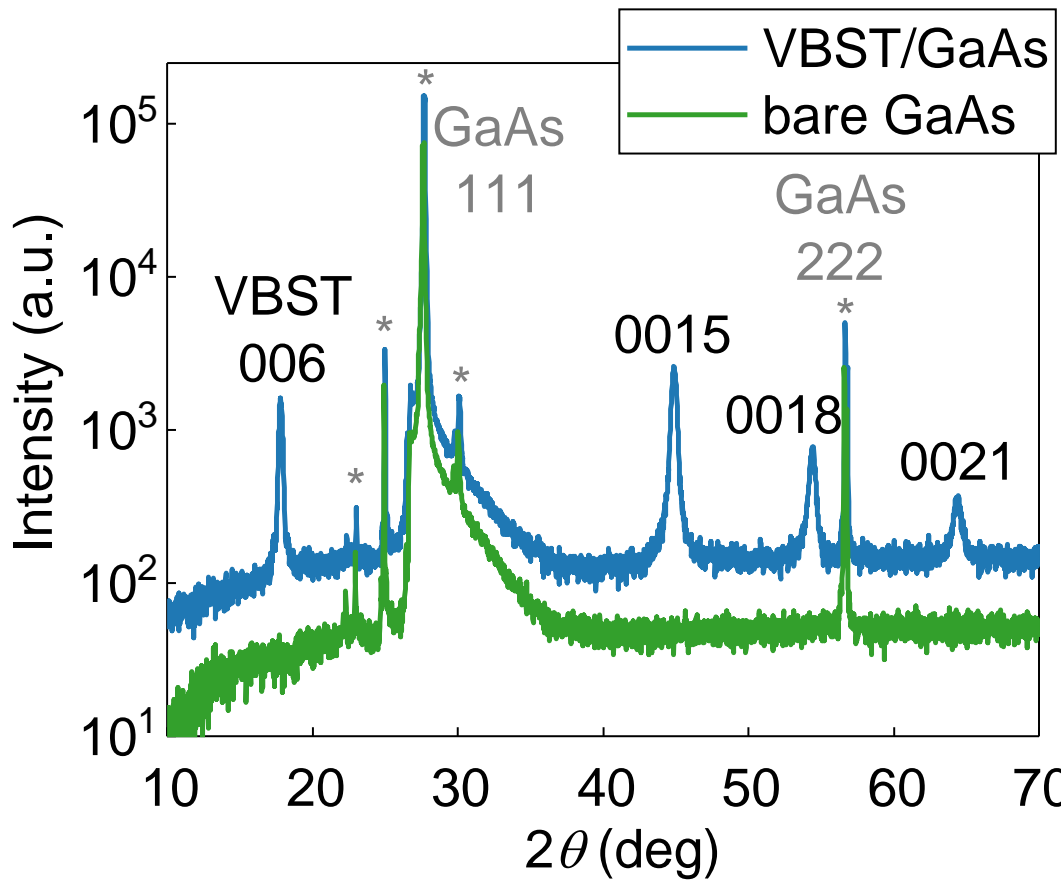

**Figure S2.** XRD $2\theta$-$\omega$ scan of VBST on GaAs and bare GaAs substrate.

The X-ray diffraction (XRD) in **Figure S2** was measured on a 29 quintuple layer (QL) $V_{0.14}(Bi_{0.22}Sb_{0.78})_{1.86}Te_3$ thin film on GaAs (111)B substrate using an X-ray powder diffractometer with Cu Kα radiation (Panalytical X'Pert Pro). The X-ray intensity is scanned with regard to $2\theta$, where $\theta$ is the angle between the incident beam and the crystallographic reflecting plane. As a reference, a bare GaAs (111)B substrate was also measured. Apart from the peaks with a * from the substrate itself, which include GaAs (111) & (222), the XRD data of VBST reveal four additional sharp peaks from VBST (006), (0015), (0018), and (0021). This proves the single crystalline phase of VBST with high quality.

## 3. Quantum Anomalous Hall Effect in VBST

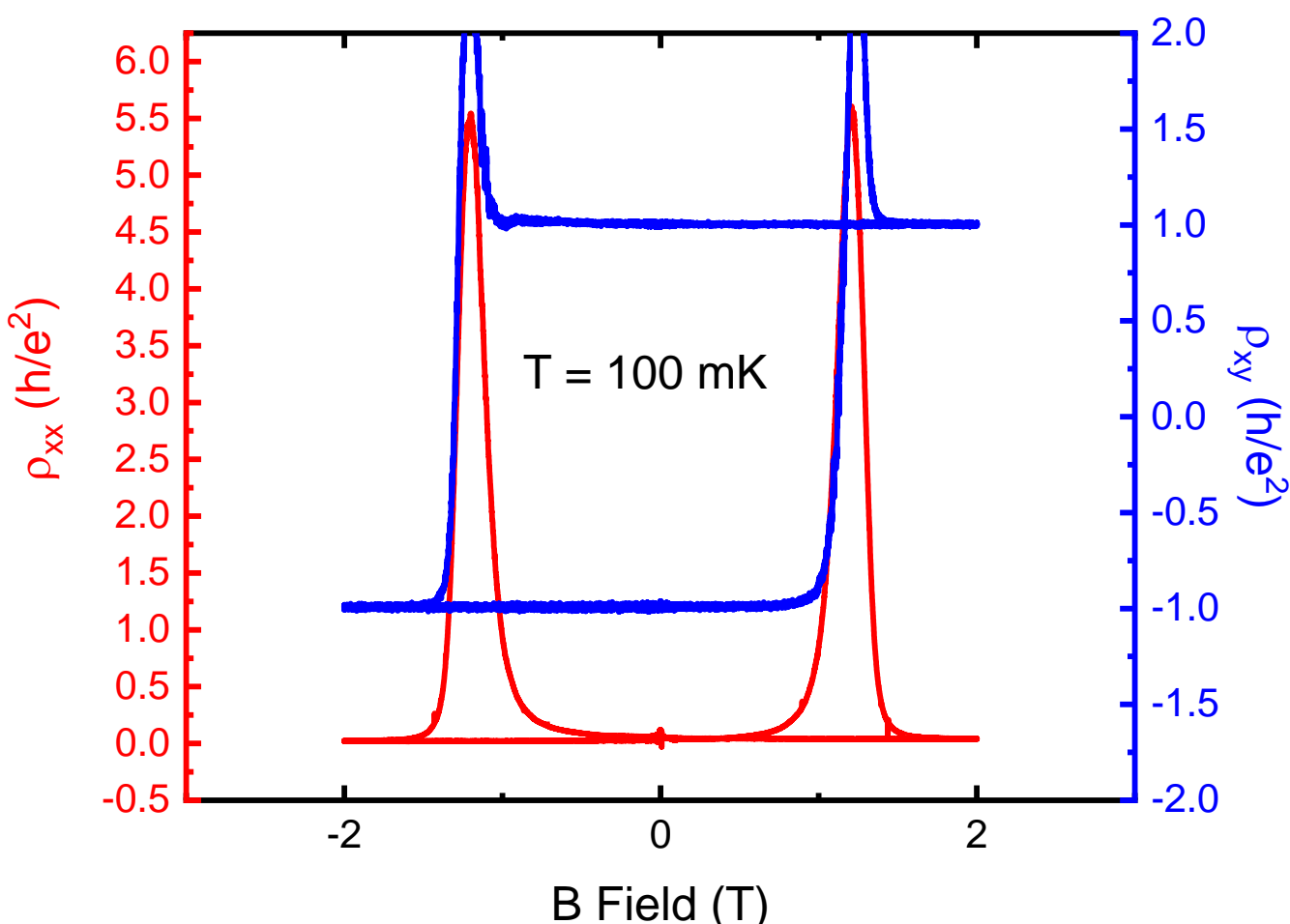

**Figure S3.** Quantum Anomalous Hall effect in VBST (Device No.3) taken under $T$=100 mK.

Device No.3 was measured inside a dilution refrigerator insert for the Quantum Design physical property measurement system (PPMS) with a base temperature of 50 mK. No gate was applied during the measurement. As shown in **Figure S3**, at $T$=100 mK, Device No.3 exhibits a perfectly quantized Hall resistivity $\rho_{xy}$ of h/e$^2$ and an almost vanishing longitudinal resistivity $\rho_{xx}$, thus confirming the presence of the quantum anomalous Hall effect. Notably, Device No.3 has an even larger coercive field of 1.3 T at $T$=100 mK.

## 4. Additional Transport Data in Different Devices

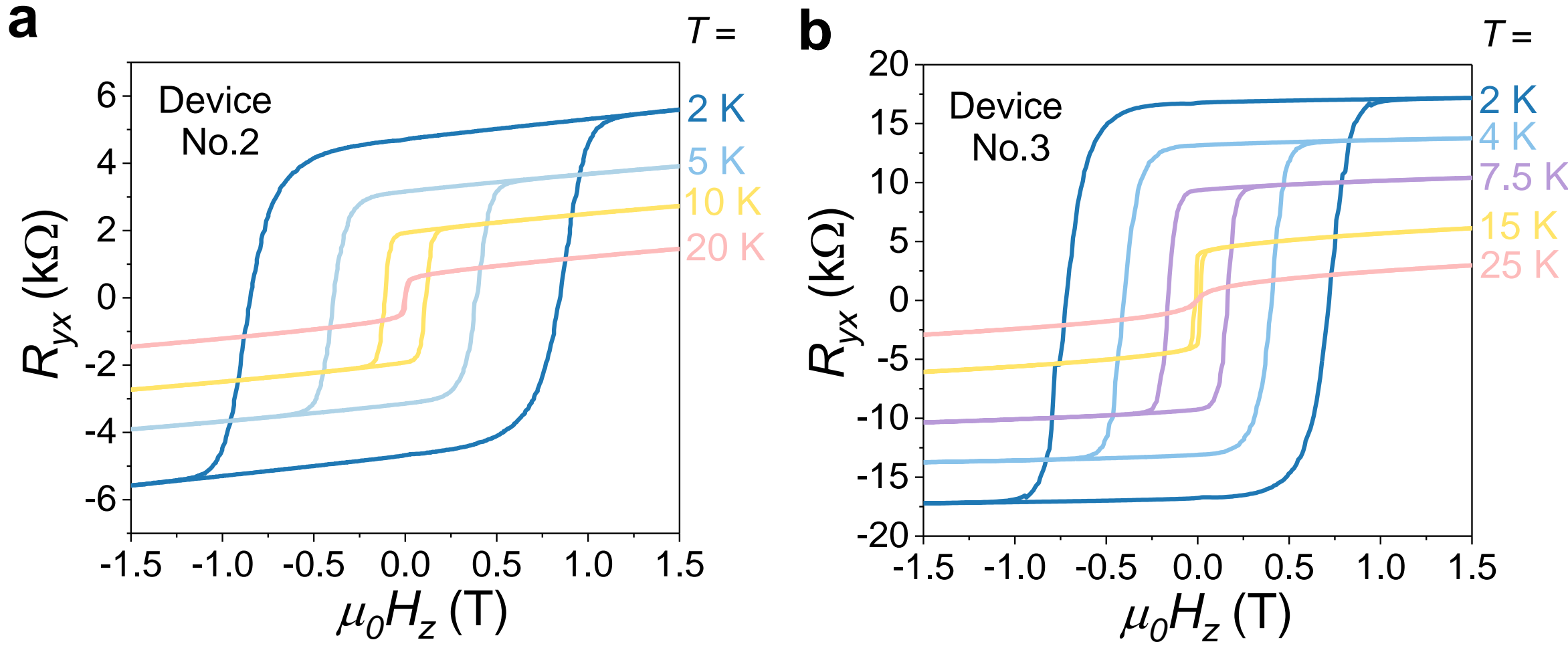

**Figure S4.** Anomalous Hall resistance $R_{yx}$ as a function of an external out-of-plane magnetic field $H_z$ at various temperatures in a) Device No.2 and b) Device No.3.

The Hall resistances of Device No.2, 6 QL $V_{0.14}(Bi_{0.18}Sb_{0.82})_{1.86}Te_3$ and Device No.3, 6 QL $V_{0.14}(Bi_{0.29}Sb_{0.71})_{1.86}Te_3$ as a function of an external out-of-plane magnetic field $H_z$ at various temperatures are presented in **Figure S4**. Device No.3 has a top gate, but no gate was applied. Anomalous Hall resistances $R_{AHE}$ of 16.8 kΩ & 4.69 kΩ and coercive fields of 0.72 T & 0.86 T are exhibited, respectively, at $T$=2 K. Notably, Device No.2 shows a large positive ordinary Hall effect as well, indicating its p-type carrier and its much lower Fermi level.

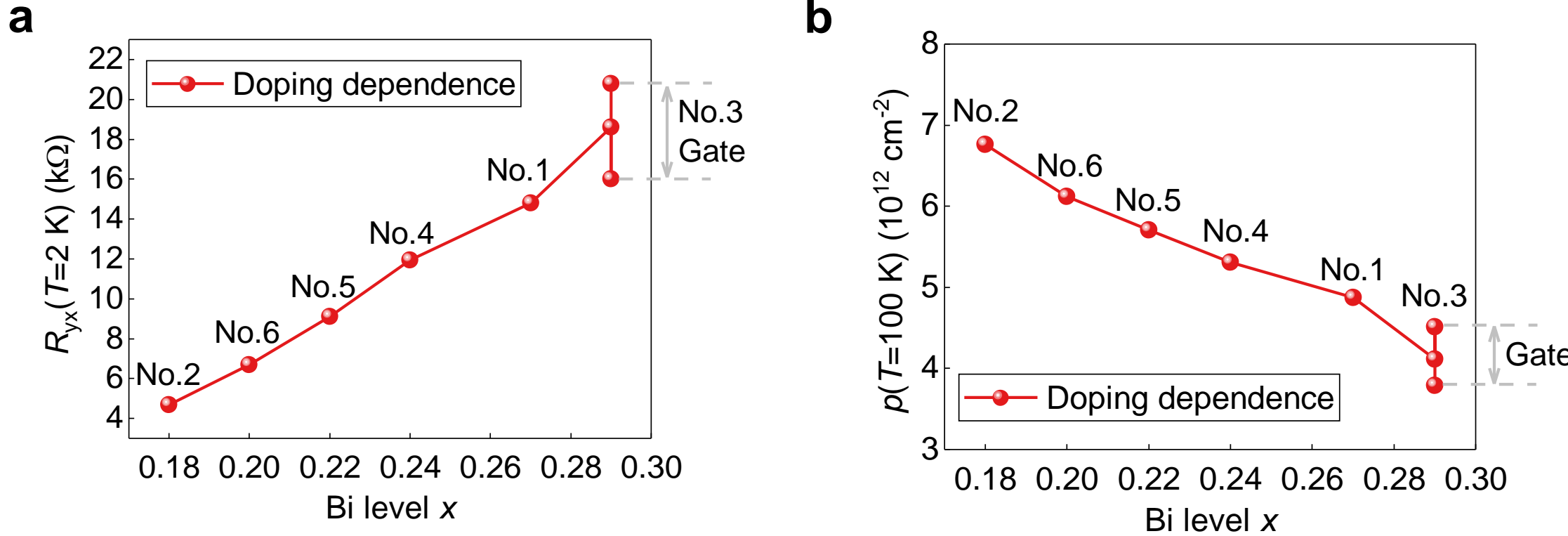

**Figure S5**. Doping dependence of a) anomalous Hall resistance $R_{yx}$ at $T$=2 K and b) hole density $p$ at 100 K across various devices.

By tuning the Bi doping level $x$ from 0.29 to 0.18, the anomalous Hall resistance $R_{yx}$ in **Figure S5** shrinks further beyond the range of electrostatic gating (as indicated by the vertical bar at $x$=0.29, where the middle dot is the zero-gate value) from 20.8 kΩ to 4.69 kΩ at $T$=2 K as the 2D hole density $p$ increases from $3.79\times10^{12}$ cm$^{-2}$ to $6.76\times10^{12}$ cm$^{-2}$ at $T$=100 K, indicating Fermi level is tuned even lower and further away from the Dirac point.

## 5. Measurement Scheme of Current-Driven Switching

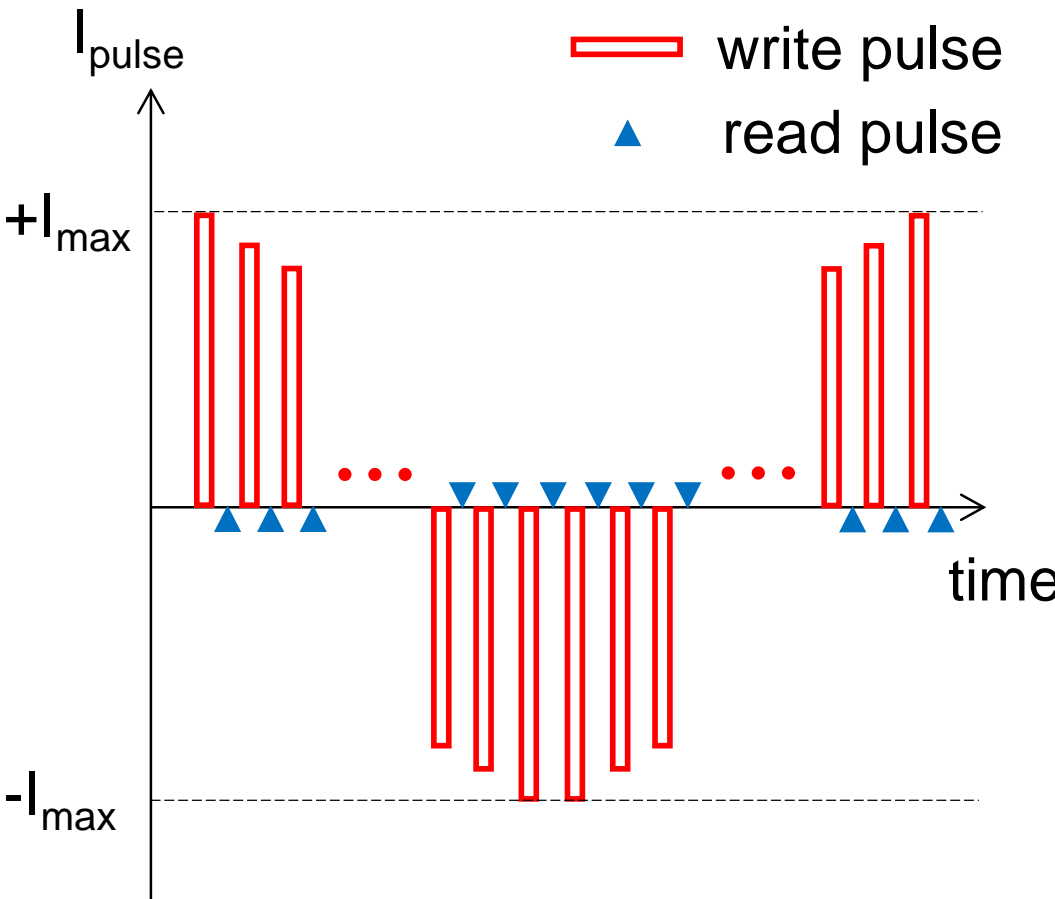

**Figure S6**. The measurement scheme of current-induced switching by sending a series of current pulses, each with a duration of 1 msec and separated by 1 sec.

Before the measurements, the magnetization was initialized by an external magnetic field to the **+z** direction. As illustrated in **Figure S6**, a series of writing current pulses, with a duration of 1 msec each, a separation of 1 sec between two, and amplitudes sweeping from $+I_{max}$ to $-I_{max}$ and then back to $+I_{max}$, were applied in order to switch the magnetization and meanwhile minimize the Joule heating effect. After each writing pulse, a reading current pulse with an amplitude of 1 μA was applied to read the Hall voltage from the device.

## 6. Current-Driven Switching under Various In-Plane Assisting Fields

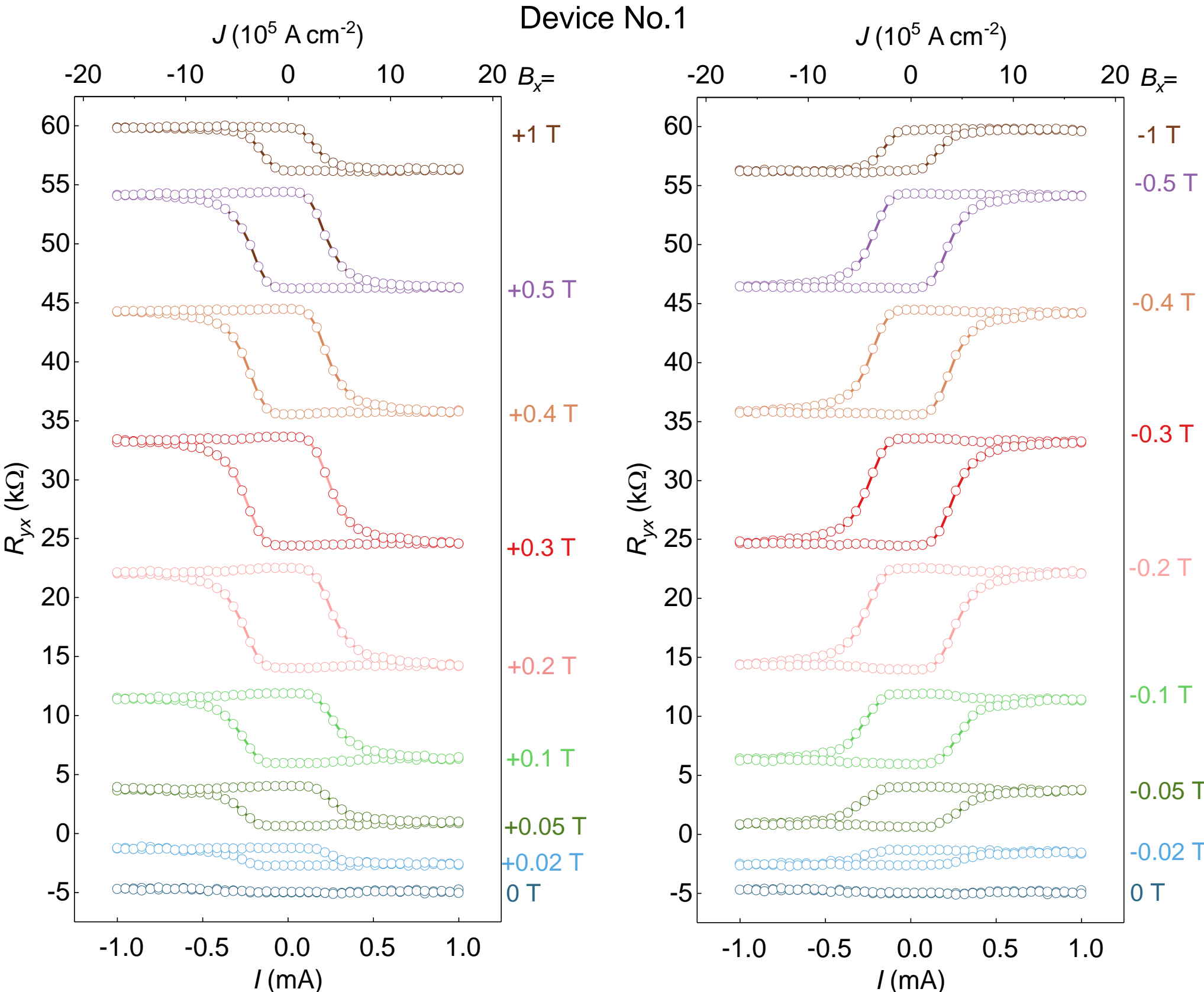

**Figure S7.** The Hall resistance $R_{yx}$ of Device No.1 after applying each current pulse, as a function of the pulse amplitude $I$ as well as the current density $J$, under various in-plane assisting magnetic fields $B_x$. All measurements were carried out at $T$=2 K.

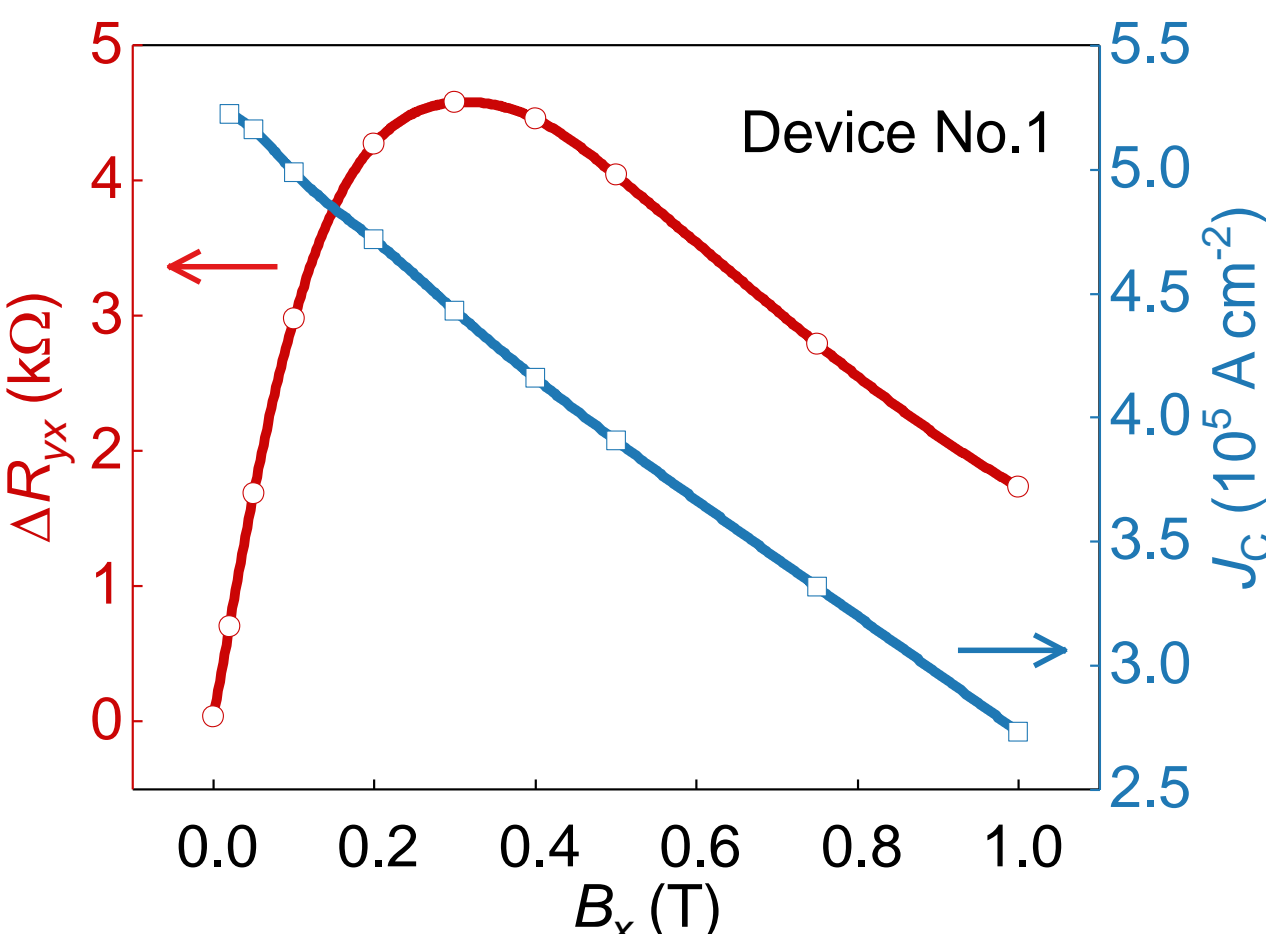

**Figure S8.** The switched portion of the Hall resistance $\Delta R_{yx}$ and the switching current density $J_C$ as a function of the in-plane assisting magnetic field $B_x$.

In order to systematically study the effect of the in-plane assisting magnetic field $B_x$ on SOT switching, the full switching loops in Device No.1 under various in-plane assisting fields $B_x$ are presented in **Figure S7**, and the switched portion of the Hall resistance $\Delta R_{yx}$ and the switching current density $J_C$ as a function of $B_x$ are presented in **Figure S8**. Since the full switching loops under a positive or negative field with the same absolute value are almost perfectly anti-symmetric, $\Delta R_{yx}$ and $J_C$ are thus calculated by averaging the two. $\Delta R_{yx}$= 9.2 kΩ is the best at $|B_x|$= 0.3 T. A smaller $|B_x|$ cannot produce enough canting of **m** towards the x direction to make the z-component of $\mathbf{B}_{DL}$ large enough, while a larger $|B_x|$ pins **m** too much towards the x direction, so in both cases $\Delta R_{yx}$ becomes smaller. For the critical switching current density $J_C$, as $|B_x|$ gets larger, both the larger z-component of $\mathbf{B}_{DL}$ and a smaller magnetization in the z direction will facilitate the switching process, making $J_C$ smaller.

## 7. Current-Driven Switching with the Largest Switching Ratio

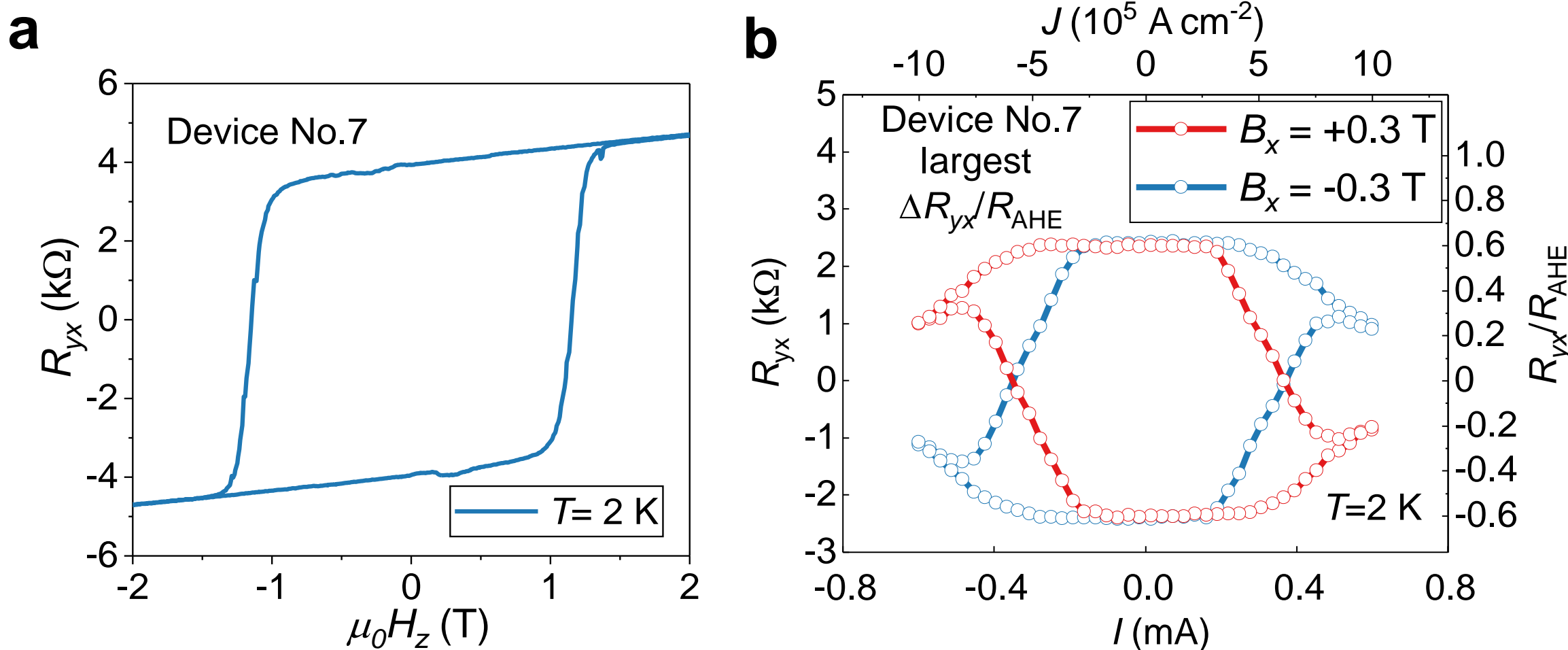

**Figure S9.** a) Anomalous Hall resistance $R_{yx}$ as a function of an external out-of-plane magnetic field $H_z$ at various temperatures in Device No.7. b) The Hall resistance $R_{yx}$ and its ratio to the anomalous Hall resistance $R_{AHE}$ after applying each writing pulse as a function of the pulse amplitude $I$ and the current density $J$, taken in Device No.7. All measurements were carried out at $T$=2 K.

The magnetic V doping level is increased in Device No.7, 6 QL $V_{0.19}(Bi_{0.20}Sb_{0.80})_{1.81}Te_3$, thus exhibiting a larger perpendicular magnetic anisotropy (PMA) as indicated by a higher coercive field of 1.15 T, as demonstrated in **Figure S9**a. Thanks to the more robust ferromagnetism, Device No.7 exhibits the largest switching ratio $\Delta R_{yx}/2R_{AHE}$ of 60% in MTI at $T$=2 K in Figure S9b, among the largest switching ratios ever measured in MTI-based SOT systems, with $\Delta R_{yx}$ = 4.72 kΩ and $R_{AHE}$= 3.94 kΩ. The switching current density is 6.0×10$^5$ A/cm$^2$, larger because its enhanced PMA creates an additional energy barrier for magnetization switching.

## 8. Derivation of the In-plane Low-Field Second-Harmonic Method

Here we present a detailed derivation of the in-plane low-field second-harmonic method for measuring the damping-like and field-like SOT effective fields.[1,2]

Before the second-harmonic measurements, the magnetization direction **m** is initialized by an external magnetic field to $\pm\mathbf{z}$ direction. During the measurements, either an external longitudinal or transverse magnetic field, $\mathbf{B}_\mathrm{L} = B_\mathrm{L}\mathbf{x}$ or $\mathbf{B}_\mathrm{T} = -B_\mathrm{T}\mathbf{y}$ is applied to determine either $\mathbf{B}_\mathrm{DL} = B_\mathrm{DL}\mathbf{m} \times \boldsymbol{\zeta} = \pm B_{DL}\mathbf{x}$ or $\mathbf{B}_\mathrm{FL} = B_\mathrm{FL}\boldsymbol{\zeta} = -B_\mathrm{FL}\mathbf{y}$, as illustrated in Figure 2a, since $\mathbf{B}_\mathrm{L} \parallel \mathbf{B}_\mathrm{DL}$ and $\mathbf{B}_\mathrm{T} \parallel \mathbf{B}_\mathrm{FL}$.

For simplicity, the case of applying $\mathbf{B}_\mathrm{L} = B_\mathrm{L}\mathbf{x}$ only when the magnetization direction $\mathbf{m} = +\mathbf{z}$ is discussed. $\mathbf{B}_\mathrm{L}$ will generate a tilting of **m** towards the **x** direction, and by the Stoner–Wohlfarth model, the tilting angle $\theta$ with respect to the **z** axis is expressed as

$$\theta = \frac{B_\mathrm{L}}{\mu_0 D} \tag{S1}$$

Here, $D = 2K/M_\mathrm{S} - 4\pi M_\mathrm{S}$ is a constant determined by saturation magnetization $M_\mathrm{S}$ and magnetic anisotropy $K$.

Then, in order to generate the SOT effective fields, a sinusoidal excitation current of $I(t) = I_0 \sin\omega t$ is applied, thus exerting an additional longitudinal field $\Delta B_\mathrm{L} = B_\mathrm{DL}\sin\omega t$ from the DL SOT effective field $\mathbf{B}_\mathrm{DL} = B_\mathrm{DL}\mathbf{x}$, which is linearly proportional to the current amplitude $I(t)$. This additional longitudinal field will further change the tilting angle $\theta$ of the magnetization with respect to the **z** axis.

$$\theta = \frac{B_\mathrm{L} + B_\mathrm{DL}\sin\omega t}{\mu_0 D} \tag{S2}$$

Since the anomalous Hall resistance $R_{yx}$ is linearly proportional to the z-component of the magnetization, assuming the tilting angle is small enough, or $\theta \ll 1$, the anomalous Hall resistance $R_{yx}$ after the tilting now becomes

$$R_{yx} = R_\mathrm{AHE}\cos\theta \approx R_0\left(1 - \frac{\theta^2}{2}\right) \tag{S3}$$

where $R_\mathrm{AHE}$ is the anomalous Hall resistance at zero field. Therefore, the Hall voltage $V_{yx} = R_{yx}I_0\sin\omega t = V_{yx}^\mathrm{DC} + V_{yx}^{1\omega}\sin\omega t + V_{yx}^{2\omega}\sin(2\omega t + \pi/2) + V_{yx}^{3\omega}\sin 3\omega t$ now has four components that are DC, first, second, and third harmonic.

Note that the first harmonic voltage $V_{yx}^{1\omega}$ is measured in phase, and the second harmonic voltage $V_{yx}^{2\omega}$ is measured $\pi/2$ out of phase with regard to the excitation current of $I(t) = I_0\sin\omega t$, and they are expressed in **Equation S4** and **S5**.

$$V_{yx}^{1\omega} = R_{\mathrm{AHE}} I_0 \left[ 1 - \frac{B_{\mathrm{L}}^2 + \frac{3}{4} B_{\mathrm{DL}}^2}{2(\mu_0 D)^2} \right] \tag{S4}$$

$$V_{yx}^{2\omega} = \frac{R_{\mathrm{AHE}} I_0 B_{\mathrm{DL}}}{2(\mu_0 D)^2} B_{\mathrm{L}} \tag{S5}$$

Therefore, the experimental results of the in-phase first harmonic voltage $V_{yx}^{1\omega}$ and the out-of-phase second harmonic voltage $V_{yx}^{2\omega}$ measured by lock-in amplifiers, as presented in Figure 3, also reveal a parabolic and linear dependence on the external longitudinal magnetic field $B_{\mathrm{L}}$, respectively. Note that when the magnetization direction $\mathbf{m} = -\mathbf{z}$, $V_{yx}^{1\omega}$ changes sign while $V_{yx}^{2\omega}$ still keeps the same positive slope because both $R_0$ and $\mathbf{B}_{\mathrm{DL}} = B_{\mathrm{DL}} \mathbf{m} \times \boldsymbol{\zeta} = -B_{\mathrm{DL}} \mathbf{x}$ changes sign with $\mathbf{m}$. Therefore, we can obtain the expression for $B_{\mathrm{DL}}$ as presented in Equation 2 in the main text

$$B_{\mathrm{DL}} = -2 \frac{\partial V_{yx}^{2\omega}}{\partial B_{\mathrm{L}}} \bigg/ \frac{\partial^2 V_{yx}^{1\omega}}{\partial B_{\mathrm{L}}^2} \tag{S6}$$

Following a similar line of arguments, simply by replacing $\mathbf{B}_{\mathrm{DL}} = B_{\mathrm{DL}} \mathbf{x}$ with $\mathbf{B}_{\mathrm{FL}} = -B_{\mathrm{FL}} \mathbf{y}$ and $\mathbf{B}_{\mathrm{L}} = B_{\mathrm{L}} \mathbf{x}$ with $\mathbf{B}_{\mathrm{T}} = -B_{\mathrm{T}} \mathbf{y}$ in the above derivation, we could obtain a similar expression for $B_{\mathrm{FL}}$ in **Equation S7**.

$$B_{\mathrm{FL}} = -2 \frac{\partial V_{yx}^{2\omega}}{\partial B_{\mathrm{T}}} \bigg/ \frac{\partial^2 V_{yx}^{1\omega}}{\partial B_{\mathrm{T}}^2} \tag{S7}$$

## 9. Field-Like Torque in Device No.2

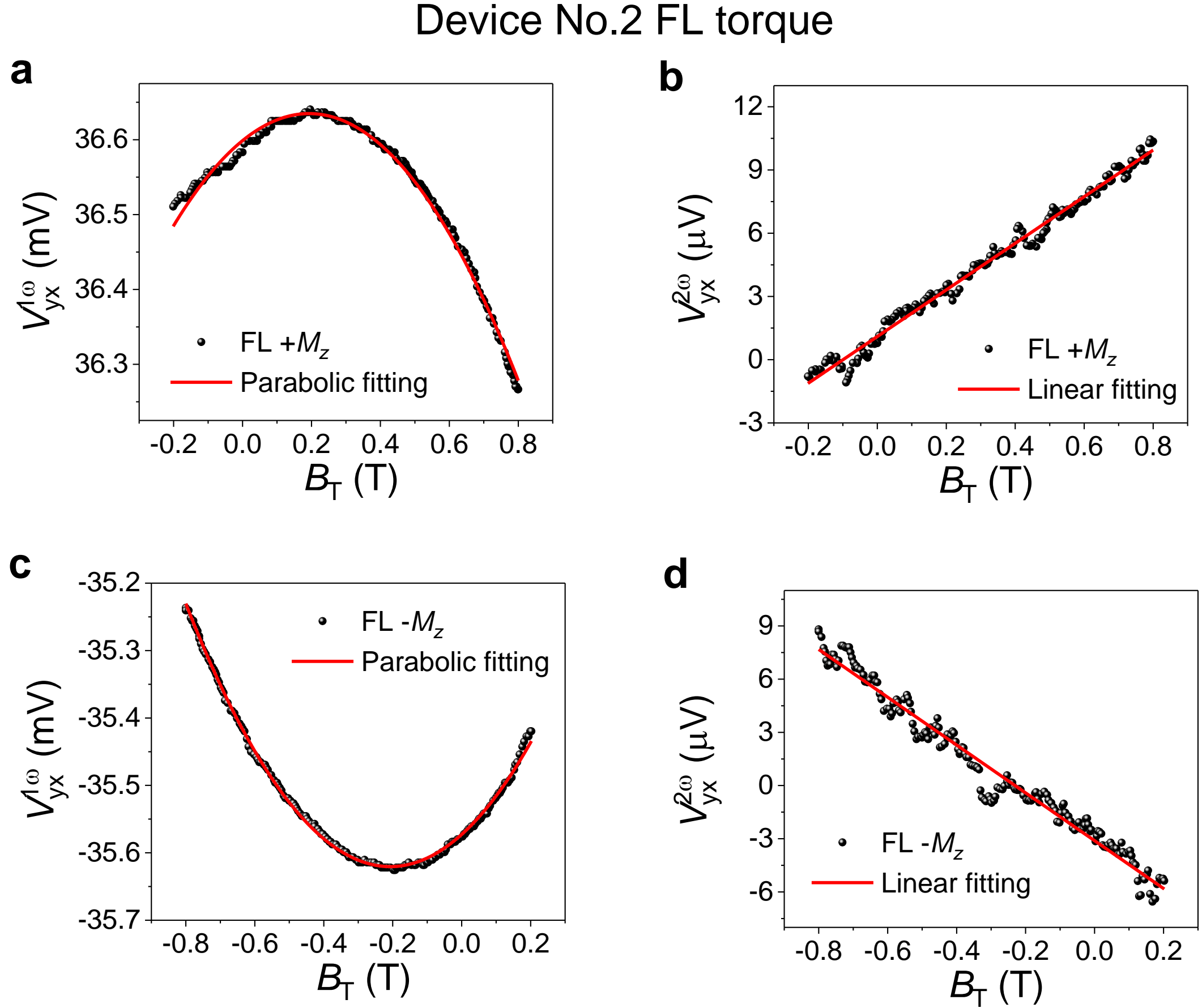

**Figure S10.** In-plane low-field second-harmonic measurements of the field-like (FL) SOT effective field in Device No.2. a) In-phase first-harmonic and b) out-of-phase second-harmonic Hall voltages $V_{yx}^{1\omega}$ and $V_{yx}^{2\omega}$ as a function of an external transverse magnetic field $B_T$ when the magnetization is along the $+\mathbf{z}$ direction (+$M_z$). c) In-phase first-harmonic and d) out-of-phase second-harmonic Hall voltages when the magnetization is along the $-\mathbf{z}$ direction (-$M_z$). All measurements were carried out at $T$=2 K by applying a sinusoidal current of 10 μA.

By sweeping an external transverse magnetic field $\mathbf{B}_\mathrm{T} = -B_\mathrm{T}\mathbf{y}$ and measuring the in-phase first harmonic voltage $V_{yx}^{1\omega}$ and the out-of-phase second harmonic voltage $V_{yx}^{2\omega}$, as presented in **Figure S10**, the field-like (FL) torque effective field of Device No.2 is also determined by Equation S7 to be $B_\mathrm{FL} = (1.178 \pm 0.004) \times 10^{-2}$ T, corresponding to $B_\mathrm{FL}/J = (7.07 \pm 0.02) \times 10^{-7}$ T A$^{-1}$ cm$^2$, about 45% of the DL torque.

## 10. Damping-Like Torque in Device No.1

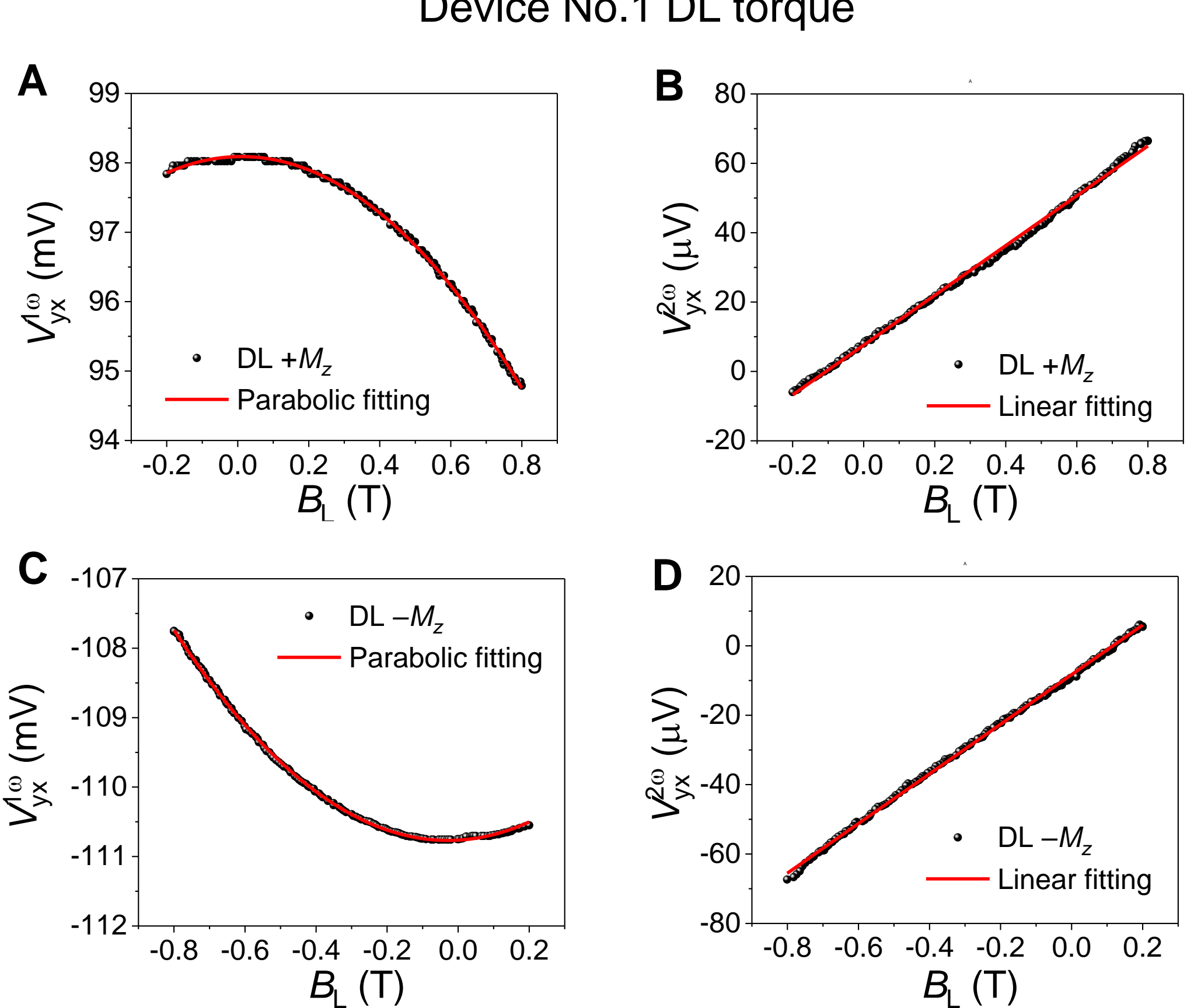

**Figure S11.** In-plane low-field second-harmonic measurements of the damping-like (DL) SOT effective field in Device No.1. a) In-phase first-harmonic and b) out-of-phase second-harmonic Hall voltages $V_{yx}^{1\omega}$ and $V_{yx}^{2\omega}$ when the magnetization is along the $+\mathbf{z}$ direction (+$M_z$). c) In-phase first-harmonic and d) out-of-phase second-harmonic Hall voltages when the magnetization is along the $-\mathbf{z}$ direction (-$M_z$). All measurements were carried out at $T$=2 K by applying a sinusoidal current of 10 μA.

In another device of Device No.1, 6 QL $V_{0.14}(Bi_{0.27}Sb_{0.73})_{1.86}Te_3$, similar second-harmonic measurements of the DL effective field was carried out, as presented in **Figure S11**, giving $B_{\mathrm{DL}} = (1.34 \pm 0.02) \times 10^{-2}$ T, $B_{\mathrm{DL}}/J = (8.0 \pm 0.3) \times 10^{-7}$ T A$^{-1}$ cm$^2$, and $q_{\mathrm{ICS}} = 2.0 \pm 0.1$ nm$^{-1}$.

## 11. Exclusion of the Nernst Effect in the Second Harmonics

Device No.8 Angle-Resolved 2nd Harmonics $B_{DL}$=2.07×10$^{-2}$ T

**Figure S12.** Angle-resolved second-harmonic measurements of the damping-like (DL) SOT effective field and the Nernst effect contribution in Device No.8. a) Normalized magnetization *m* as a function of the in-plane external field *H* to determine the anisotropy field $H_K$. b) Second-harmonic resistance $R_{yx}^{2\omega}$ as a function of the angle $\theta_H$ of the external field $H_{ext}$ with regards to the *z*-axis. c) Second-harmonic resistance $R_{yx}^{2\omega}$ as a function of angle $\theta$ of the magnetization *M* with regards to the *z*-axis. d) The linear dependence of $dR_{yx}^{2\omega}/d(\sin\theta)$ on $1/(H_{ext}+H_K)$, where the SOT and Nernst contributions are separated into the slope and the intercept, respectively. All measurements were carried out at *T*=2 K by applying a sinusoidal current of 10 μA.

In this paper, second harmonic measurements are conducted by applying a small in-plane external field $B_L$ to characterize the SOT damping-like effective field $B_{DL}$. However, this method is susceptible to thermoelectric effect contributions, especially the anomalous Nernst effect (ANE) from an out-of-plane (OOP) thermal gradient produced by Joule heating. The ANE and SOT coexist and give rise to AC transverse signals with comparable symmetry, thus adding a Nernst term to Equation S5 for the second harmonic voltage $V_{yx}^{2\omega}$, as shown in **Equation S8**

$$V_{yx}^{2\omega} = \frac{R_{\mathrm{AHE}} I_0 B_{\mathrm{DL}}}{2(\mu_0 D)^2} B_{\mathrm{L}} + I_0 R_{\mathrm{Nernst}} \sin\theta = \frac{R_{\mathrm{AHE}} I_0 B_{\mathrm{DL}}}{2(\mu_0 D)^2} B_{\mathrm{L}} + I_0 R_{\mathrm{Nernst}} \frac{B_{\mathrm{L}}}{\mu_0 D} \quad \text{(S8)}$$

Here, in the second Nernst term, $R_{\mathrm{Nernst}} \propto I_0 \alpha \nabla T_{\mathrm{OOP}}$ is the resistance from the ANE contribution, where $\alpha$ is the ANE coefficient and $\nabla T_{\mathrm{OOP}}$ is the out-of-plane (OOP) temperature gradient. $R_{\mathrm{AHE}}$ is the anomalous Hall resistance at zero field, $I_0$ is the AC current amplitude, and $D = 2K/M_{\mathrm{S}} - 4\pi M_{\mathrm{S}}$ is a constant determined by saturation magnetization $M_{\mathrm{S}}$ and magnetic anisotropy $K$. Both the SOT and the Nernst terms in the second harmonic voltage are linearly dependent on the external in-plane field $B_{\mathrm{L}}$, making it hard to disentangle the two.[3]

In order to separate the Nernst contribution, a different approach of angle-resolved second harmonics is adopted. An external field $H_{\mathrm{ext}}$ is applied with a fixed amplitude and the angle $\theta_{\mathrm{H}}$ of the external field with regards to the $z$-axis is swept, as in the schematic shown in **Figure S12**b. As given by **Equation S9**, the second harmonic resistance $R_{yx}^{2\omega}$ has the following dependence on the tilting angle $\theta$ of the magnetization with respect to the $z$-axis, as in the schematic shown in Figure S12c.

$$R_{yx}^{2\omega} = \frac{1}{2} I_0 R_{\mathrm{AHE}} \frac{B_{\mathrm{DL}}}{\mu_0 (H_{\mathrm{ext}} + H_K)} \sin\theta + R_{\mathrm{Nernst}} \sin\theta \quad \text{(S9)}$$

Here, $H_K$ is the anisotropy field. $R_{yx}^{2\omega}$ is linearly proportional to $\sin\theta$ and the slope is given by **Equation S10**

$$\frac{dR_{yx}^{2\omega}}{d(\sin\theta)} = \frac{1}{2} I_0 R_{\mathrm{AHE}} \frac{B_{\mathrm{DL}}}{\mu_0 (H_{\mathrm{ext}} + H_K)} + R_{\mathrm{Nernst}} \quad \text{(S10)}$$

Therefore, by analyzing the linear dependence of $dR_{yx}^{2\omega}/d(\sin\theta)$ on $1/(H_{\mathrm{ext}} + H_K)$, the SOT and Nernst contributions are separated into the slope and the intercept, respectively. Note that the Spin Seeback Effect (SSE) has a similar symmetry and contributes similarly to the $R_{\mathrm{Nernst}}$ term, so this method can be used to exclude the SSE as well.[4]

To estimate and exclude the error of the SOT effective field caused by the Nernst effect in the in-plane low-field second harmonic measurements, we fabricated a new VBST Device No.8, 6 QL $V_{0.14}(Bi_{0.19}Sb_{0.81})_{1.86}Te_3$ and applied both the angle-resolved and in-plane low-field second harmonic methods on this same device to compare the results, with the same excitation current of $I_0 = 10$ μA.

To determine the anisotropy field $H_{\mathrm{K}}$, in Figure S12a, the normalized magnetization curves $m = M_{//}/M_{\mathrm{S}}$ (where $M_{//}$ is the magnetization component along the field direction and $M_{\mathrm{S}}$ is the saturation magnetization) are presented as a function of the external field $H$ along either the hard or easy axis. As a hard ferromagnet, VBST has a strong perpendicular magnetic anisotropy (PMA) and does not require any field to drive the magnetization to the easy axis ($z$), therefore the easy-axis magnetization curve is $H$=0 and $m$=1 as in the dashed line. The hard-axis magnetization curve is obtained by applying an in-plane $H$ field, measuring the Hall resistance $R_{yx} \propto m_z$, and calculating the in-plane $m = \sqrt{1 - m_z^2} = \sqrt{1 - \left(R_{yx}/R_{\mathrm{AHE}}\right)^2}$, where $R_{\mathrm{AHE}}$ is the saturated AHE resistance. Since the magnetic anisotropy constant $K = H_K M_{\mathrm{S}}/2 = \int H \cdot dM_{//}$, the anisotropy field $\mu_0 H_K = 2 \int \mu_0 H \cdot \mathrm{d}m = 3.81$ T is given by integrating the blue region.

The results of angle-resolved second harmonic measurements are presented in Figure S12b, c, and d. As the original data in Figure S12b, the second harmonic resistance $R_{yx}^{2\omega}$ is shown as a function of the angle $\theta_{\mathrm{H}}$ of the external field $H_{\mathrm{ext}}$ with regards to the $z$-axis, under various external fields of $\mu_0 H_{\mathrm{ext}}$=1.2 T, 1.3 T, 1.4 T, ... 2.0 T (only 1.2 T, 1.6 T, and 2.0 T are shown here for simplicity). In order to convert the $\theta_{\mathrm{H}}$ to the tilting angle $\theta$ of the magnetization with respect to the $z$-axis, the Stoner-Wohlfarth Model is used by **Equation S11**

$$-\frac{1}{2}\sin 2\theta + h\sin(\theta_H - \theta) = 0 \qquad \text{(S11)}$$

Here $h = H_{\mathrm{ext}}/H_K$ is a dimensionless coefficient. After the conversion, the second harmonic resistance $R_{yx}^{2\omega}$ as a function of $\sin\theta$ is presented in Figure S12c, showing a linear dependence under various external fields. By linear regression and obtaining the slopes, $dR_{yx}^{2\omega}/d(\sin\theta)$ is plotted as a function of $1/(H_{\mathrm{ext}} + H_K)$ in Figure S12d, also bearing a linear relationship. Using Equation S10, the SOT damping-like (DL) effective field is extracted to be $B_{\mathrm{DL}} = (2.07 \pm 0.08) \times 10^{-2}$ T from the slope of the linear fitting. The Nernst contribution $R_{\mathrm{Nernst}} = 0.86 \pm 0.29\ \Omega$ is also determined from the intercept, which is one order of magnitude smaller than the first term of SOT contributions in Equation S10.

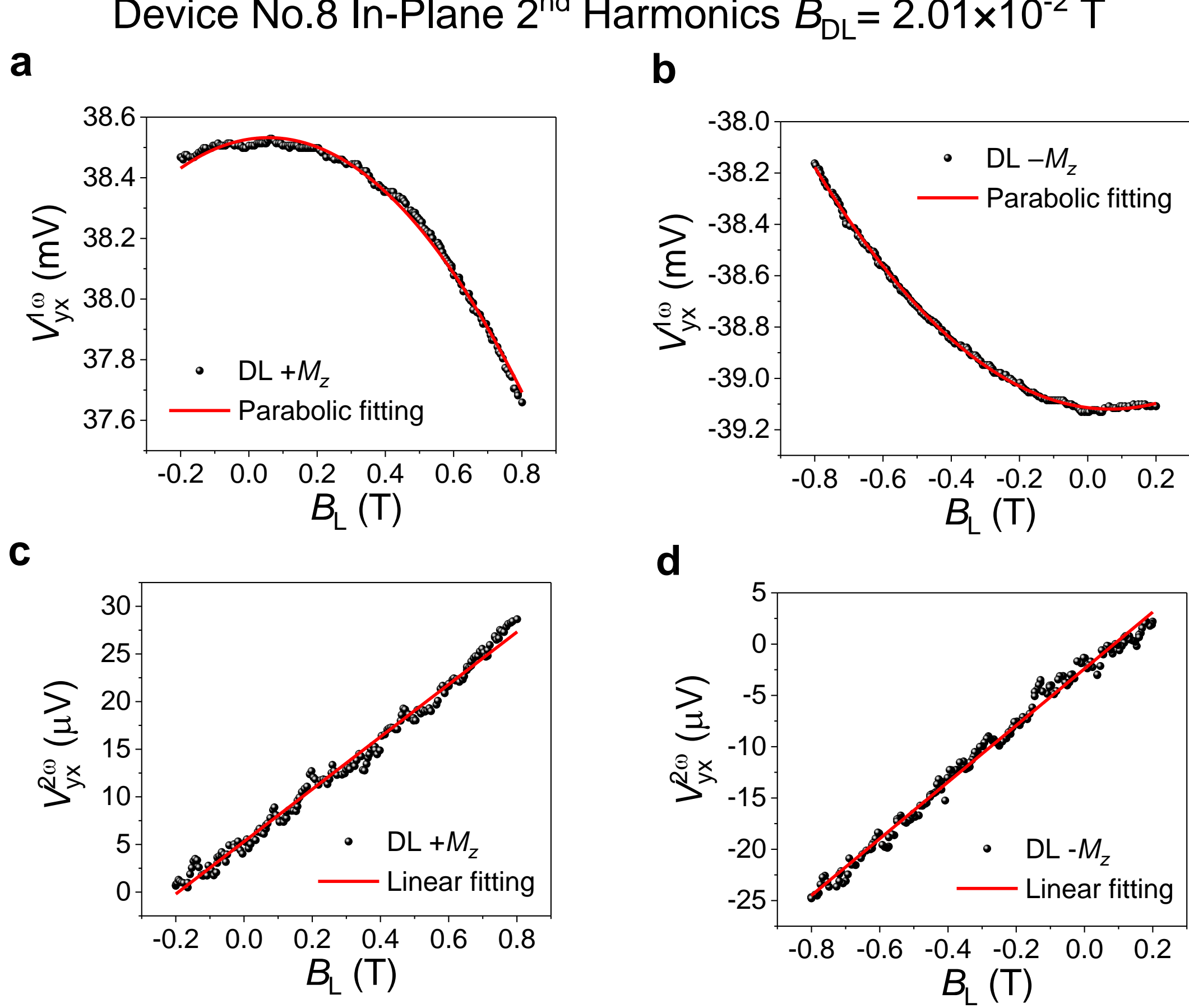

**Figure S13.** In-plane low-field second-harmonic measurements of the damping-like (DL) SOT effective field and Nernst effect contribution in Device No.8. a) In-phase first-harmonic and b) out-of-phase second-harmonic Hall voltages $V_{yx}^{1\omega}$ and $V_{yx}^{2\omega}$ when the magnetization is along the $+\mathbf{z}$ direction (+$M_z$). c) In-phase first-harmonic and d) out-of-phase second-harmonic Hall voltages when the magnetization is along the $-\mathbf{z}$ direction (-$M_z$). All measurements were carried out at $T$=2 K by applying a sinusoidal current of 10 μA.

As a comparison, the results of in-plane low-field second harmonic measurements, which are subject to the Nernst effect in Equation S8 and used in the major part of this paper, are also presented in **Figure S13**. The SOT damping-like (DL) effective field with the error from Nernst contributions is determined to be $B_{\mathrm{DL}} = (2.01 \pm 0.15) \times 10^{-2}$ T. Therefore, the error caused by the Nernst effect ($0.06 \times 10^{-2}$ T) is more than one order of magnitude smaller than the SOT effective field and smaller than the measurement error as well, thus confirming the validity of our in-plane low-field second harmonic measurements in the main text.

## 12. Current-Driven Switching Endurance

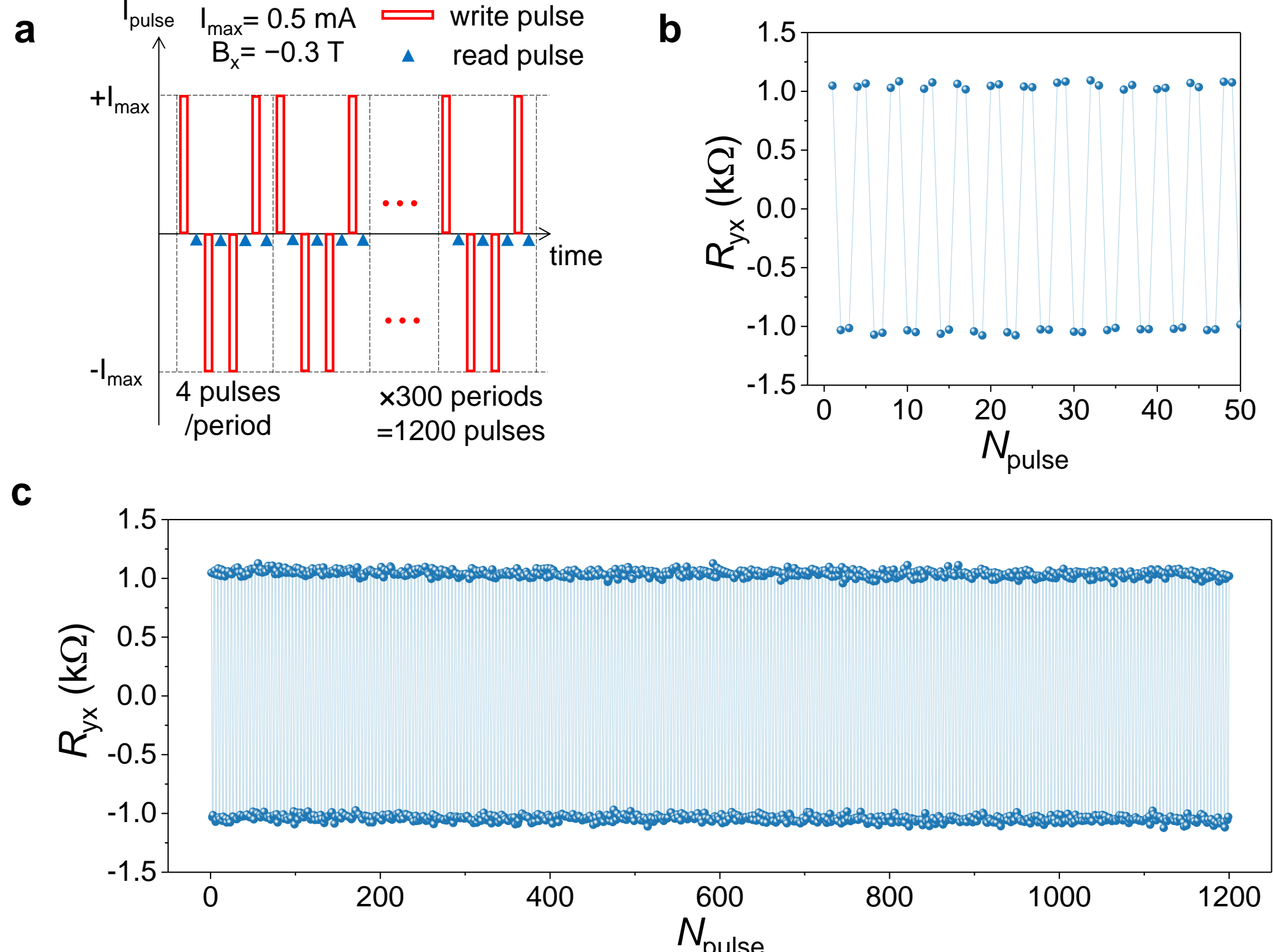

**Figure S14.** a) The measurement scheme of the current-induced switching endurance test by sending 1200 current pulses grouped by 4 pulses per period, with the same amplitude of 0.5 mA. Each pulse has a duration of 1 msec and is separated by 1 sec. b, c) The Hall resistance $R_{yx}$ after applying each writing pulse as a function of the number of write pulses $N_{\text{pulse}}$, taken in Device No.8 for the first 50 pulses in (b) and for the total 1200 pulses in (c). All measurements were carried out at $T$=2 K and $B_x$=-0.3 T.

In order to test the endurance of the current-driven switching in VBST, a switching endurance test was carried out by sending 1200 writing current pulses, as illustrated in **Figure S14**a in the VBST Device No.8, 6 QL $V_{0.14}(Bi_{0.19}Sb_{0.81})_{1.86}Te_3$. The Hall resistance data $R_{yx}$ after applying each write pulse as a function of the number of write pulses $N_{\text{pulse}}$ are presented for the first 50 pulses in Figure S14b and for the total 1200 pulses in Figure S14c. The data show minimal change in the switching amplitude, thus indicating excellent endurance over a large number of current pulses.

## 13. Temperature Dependence of Current-Driven Switching and SOT Efficiency

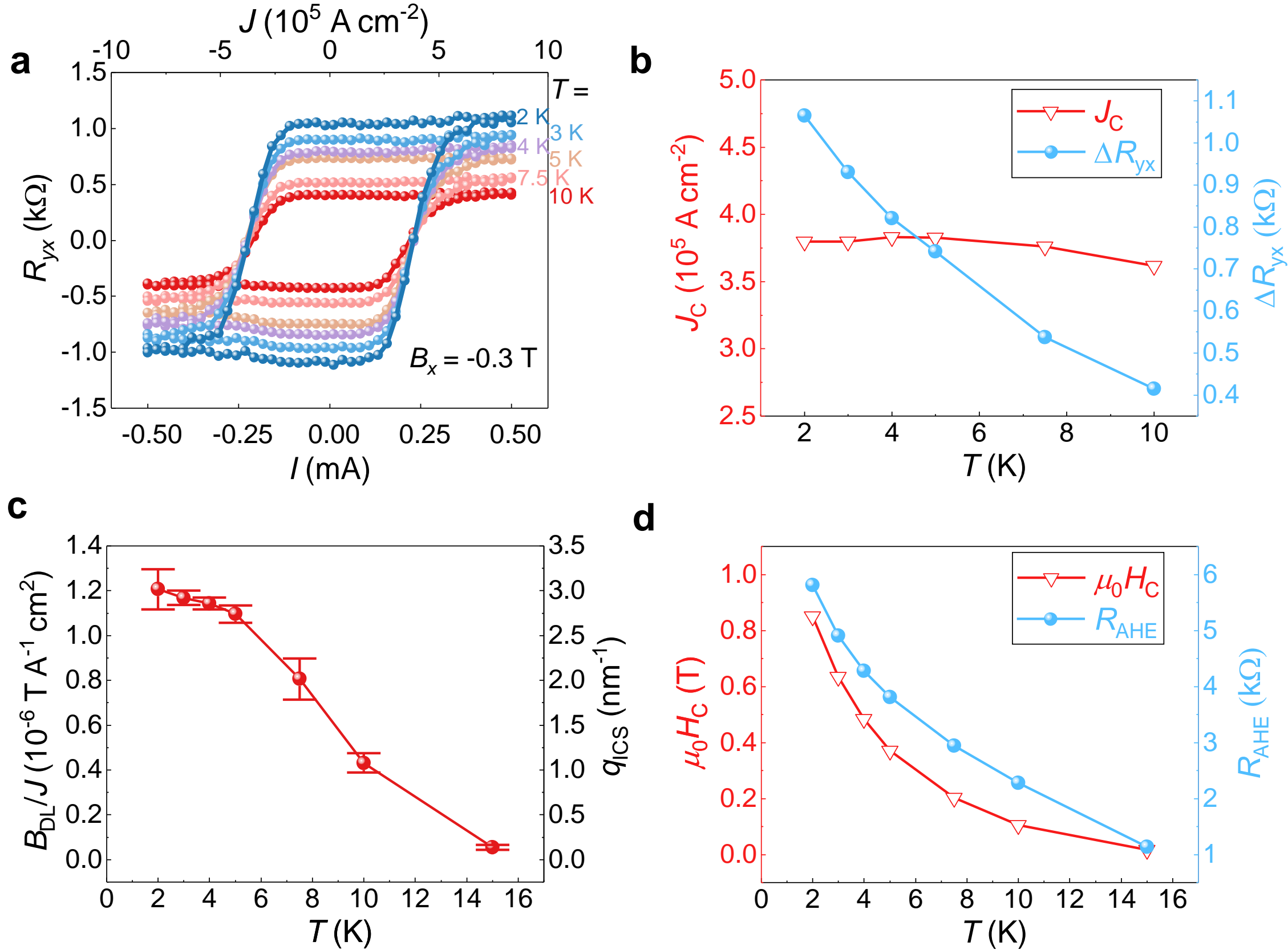

**Figure S15.** a) The Hall resistance $R_{yx}$ after applying each writing pulse as a function of the pulse amplitude $I$ as well as the current density $J$ with an in-plane assisting magnetic field $B_x$ of -0.3 T under various temperatures. b) Temperature dependence of the switching current density $J_C$ and the switched Hall resistance $\Delta R_{yx}$. c) Temperature dependence of the SOT efficiency, as indicated by the damping-like effective field to current density ratio $B_{DL}/J$ and the interfacial charge-to-spin conversion efficiency $q_{ICS}$. d) Temperature dependence of the coercive field $H_C$ and the anomalous Hall resistance $R_{AHE}$. All measurements were carried out in Device No.8.

The temperature dependence of current-driven switching and SOT efficiency in the VBST Device No.8, 6 QL $V_{0.14}(Bi_{0.19}Sb_{0.81})_{1.86}Te_3$ is presented in **Figure S15**. In Figure S15a, the current-driven switching curves under various temperatures are presented, showing well-defined switching behavior from 2 K up to 10 K. The switching current density $J_C$, as shown in Figure S15b, stays almost constant across various temperatures, with only a slight downturn above 5 K, and the switched Hall resistance $\Delta R_{yx}$ exhibits a steady decrease with increasing temperature. The temperature dependence of the SOT efficiency from the second-harmonic measurements in Figure S15c exhibits a slow decrease from 2 K to 5 K, which could be attributed to a lower spin polarization ratio of the carriers at higher temperature, and then shows a sharp downturn above 5 K, reaching almost zero at 15 K, which could be a measurement error caused by the weakened ferromagnetic order (the second-harmonic measurements assume small rotation angle of the magnetization and thus require a strong magnetic anisotropy). As a reference, the temperature dependence of the coercive field $H_C$ and the anomalous Hall resistance $R_{AHE}$ are shown in Figure S15d. The decrease of $R_{AHE}$ explains the shrinking switched Hall resistance $\Delta R_{yx}$, while a combination of decreasing $H_C$ and SOT efficiency might lead to the almost constant switching current density $J_C$.